\documentclass[journal]{IEEEtran}

\usepackage{amsmath,amsfonts,graphics,amssymb,epsfig,subfigure,color,cite}
\usepackage{amsthm,textcomp}
\theoremstyle{plain}
\newtheoremstyle{mystyle}
  {0mm}
  {0mm}
  {}
  {4mm}
  {\bfseries}
  {:}
  { }
  {\thmname{#1}\thmnumber{ #2}\thmnote{ (#3)}}
\theoremstyle{mystyle}
\usepackage{bm}

\usepackage{array}
\usepackage{multirow}
\usepackage{enumerate}
\usepackage{algorithm}
\usepackage{algpseudocode}
\usepackage{algcompatible}
\algblockdefx{WHILE}{ENDWHILE}[1]%
  {\textbf{while }#1 \textbf{do}}%
  {\textbf{end}}
\algblockdefx{FORP}{ENDFORP}[1]%
  {\textbf{for }#1 \textbf{do in parallel}}%
  {\textbf{end}}
\algblockdefx{FOR}{ENDFOR}[1]%
  {\textbf{for }#1 \textbf{do}}%
  {\textbf{end}}
\algblockdefx{noEnd}{noEndEnd}[1]
    {#1}
    {{\Statex}}
\algblockdefx{BULLET}{EndBULLET}{$\bullet$ }{}
\algnotext{EndBULLET}
\algblockdefx{PROC}{ENDPROC}[1]
    {\textbf{procedure }#1 {}}
    {\textbf{end procedure}}
\algblockdefx{IF}{ENDIF}[1]%
  {\textbf{if }#1 \textbf{then}}%
  {\textbf{end}}
\algnewcommand\algorithmicprocedure{\textbf{procedure}}
\algnewcommand\FUNC{\item[\algorithmicprocedure]}%
\algnewcommand\algorithmicendprocedure{\textbf{end procedure}}
\algnewcommand\ENDFUNC{\item[\algorithmicendprocedure]}%
 \usepackage{setspace}
\let\Algorithm\algorithm
\renewcommand\algorithm[1][]{\Algorithm[#1]\setstretch{1.4}}
\usepackage{float}
\usepackage{makecell}
\usepackage{graphicx}
\usepackage{xcolor}
\usepackage{soul}

\newtheorem{prop}{Proposition}

\usepackage{bbm}

\makeatletter
\newcommand{\vast}{\bBigg@{4.5}}
\newcommand{\Vast}{\bBigg@{7.5}}
\makeatother

\allowdisplaybreaks

\begin{document}
    \title{Physical Layer Security for Digital Semantic Communication: A Bit-Level Perspective}
	\author{Seonjung Kim, 
    \IEEEmembership{Student Member,~IEEE}, Jaehyeok Kim, and Yo-Seb Jeon, \IEEEmembership{Member,~IEEE}
	    \thanks{Seonjung Kim, Jaehyeok Kim, and Yo-Seb Jeon are with the Department of Electrical Engineering, POSTECH, Pohang, Gyeongbuk 37673, Republic of Korea (e-mails: \{seonjung.kim, jaehyeok.kim, yoseb.jeon\}@postech.ac.kr).}
	}
	\vspace{-2mm}	
	
	\maketitle
	\vspace{-12mm}

	\begin{abstract} 
    Digital semantic communication requires reliable recovery of task-relevant information at a legitimate receiver while limiting source reconstruction by an eavesdropper. Uniform bit protection, however, does not account for the heterogeneous reliability and secrecy requirements of semantic bits. This paper proposes a bit-level physical layer security (PLS) framework that learns these requirements and incorporates them into physical-layer resource allocation. We first introduce a wiretap semantic feature channel whose two-stage training learns per-bit target bit-flip probabilities for the legitimate receiver (Bob) and the eavesdropper (Eve). These targets characterize the reliability requirements at Bob and the secrecy protection requirements against Eve.
    Based on the learned targets, we formulate a joint power and artificial-noise (AN) allocation problem for a multiple-input single-output wiretap channel under a worst-case angular-separation model. For a given target profile, the optimal information--AN power split admits a closed-form solution, and the per-bit power allocation is obtained by water-filling. We further develop a leakage-aware selection method among multiple learned-target profiles and extend the design to multiple-input multiple-output channels. Simulation results demonstrate that the proposed framework achieves a better trade-off between preserving reconstruction quality at Bob and degrading that at Eve than uniform power allocation and the evaluated learning-based secure semantic
    communication schemes.
    \end{abstract}


    \section{Introduction}
    
    Semantic communication is an emerging paradigm that focuses on conveying the meaning or task-relevant content of a source, rather than faithfully reproducing every raw data bit as in traditional communication systems \cite{luo2022semantic, yang2022semantic, shao2024theory, chaccour2024less}. A representative approach is deep joint source-channel coding (JSCC), in which neural networks at the transmitter and receiver are trained in an end-to-end manner \cite{bourtsoulatze2019deep, wu2024deep, bian2025process}. Digital semantic communication quantizes the encoder output into semantic bits so that they can be transmitted using digital modulation \cite{DeepJSCC_Q, oh2025digital}. Owing to the broadcast nature of wireless channels, an eavesdropper (Eve) can overhear the semantic bitstream transmitted to a legitimate receiver (Bob) without being detected \cite{csiszar1978broadcast}. Since a semantic encoder compresses task-relevant information into a compact bitstream, the leakage of even a small number of semantic bits may reveal substantial source content \cite{meng2025survey, li2024secure}. Upper-layer cryptographic techniques can provide confidentiality, but they typically rely on computational hardness assumptions and incur processing overhead.

    Physical layer security (PLS) exploits the characteristics of wireless channels and can provide information-theoretic security without relying on Eve's limited computing power or secret-key distribution \cite{wyner1975wire, zhou2016physical}. It can protect the transmitted bits through physical-layer resources such as transmit power and artificial noise (AN). In this paper, we study PLS for digital semantic communication from a bit-level perspective: we learn how reliably each semantic bit should be delivered to Bob and how strongly it should be protected against Eve, and we use these bit-level targets to guide physical-layer resource allocation.

    \subsection{Related Work}

     In digital semantic communication, errors on different semantic bits affect task performance differently, and several recent studies have exploited this property at the bit level. In \cite{kim2026unequal}, the bit-flip probabilities learned for semantic bits were shown to differ substantially across bits, and unequal error protection (UEP) was developed to provide each bit with the reliability it requires. The semantic feature (SF) channel framework in \cite{oh2026semantic} characterizes these reliability requirements by learning the bit-flip probabilities of a binary symmetric channel (BSC) abstraction jointly with the semantic encoder and decoder. However, these bit-level approaches consider only a legitimate receiver and do not address eavesdropping.

    To protect semantic communication against eavesdropping, recent studies have incorporated PLS techniques. A widely adopted PLS technique is multiple-input single-output (MISO) beamforming combined with AN, in which AN is transmitted in the null space of Bob's channel to degrade Eve's reception \cite{goel2008guaranteeing}. Building on such techniques, AN-assisted secure semantic transmission \cite{yin2025semantic}, beamforming-based secure semantic communication \cite{cao2025physical}, and reconfigurable intelligent surface (RIS)-assisted secure semantic transmission \cite{amiriara2025physical} have been studied. Although these studies show that PLS can reduce semantic information leakage, they optimized aggregate security metrics, such as the secrecy rate or the overall reconstruction distortion at Eve, and thus did not distinguish the protection requirements of individual semantic bits.

    Another line of research trains semantic encoders against adversarial networks over wiretap channels. In \cite{marchioro2020adversarial}, a deterministic encoder was trained against an adversarial classifier to enable Bob to reconstruct images while limiting Eve's ability to infer their classes. In \cite{erdemir2022privacy}, a stochastic encoder was trained over BSCs to enable Bob to reconstruct images while limiting Eve's ability to infer their sensitive attributes. In these approaches, however, the protection was built into the encoder for the channel conditions assumed during training and is not specified at the bit level. Moreover, they did not provide a mechanism for adjusting physical-layer resources to the actual channel conditions at transmission time.

    As discussed above, existing bit-level approaches characterize the reliability requirements of semantic bits but do not consider eavesdropping, whereas existing secure semantic communication schemes do not determine how strongly each semantic bit should be protected against Eve. Consequently, none of these studies jointly characterizes the reliability and secrecy requirements of individual semantic bits and uses them to adapt physical-layer resources to the channel condition.

    \subsection{Contributions}

    To bridge this gap, we propose a bit-level PLS framework for digital semantic communication. The proposed framework learns a pair of target bit-flip probabilities for Bob and Eve for each semantic bit and uses these targets to guide information and AN power allocation at the transmitter (Alice). We develop this allocation for a MISO wiretap channel under a worst-case model of Eve's channel, without requiring precise knowledge of Eve's channel state information (CSI), and extend the design to multiple-input multiple-output (MIMO) wiretap channels at the stream level. Simulation results show that the proposed framework achieves a better trade-off between preserving the reconstruction quality at Bob and degrading that at Eve than uniform power allocation and the evaluated learning-based secure semantic communication schemes. The main contributions of this paper are summarized as follows: 

 
    \begin{itemize}
        \item  We propose a wiretap SF channel that extends the SF channel framework \cite{oh2026semantic} to a wiretap setting by modeling the end-to-end transfer of each semantic bit to Bob and Eve with independent BSCs. Based on this channel, we develop a two-stage training procedure. In the first stage, the semantic encoder, decoder, and bit-flip probabilities of Bob's SF channel are jointly trained to minimize the reconstruction distortion at Bob under a limited information budget. The resulting bit-flip probabilities serve as the bit-level reliability requirements at Bob. In the second stage, the encoder and decoder are fixed, and the bit-flip probabilities of Eve's SF channel are trained to maximize the reconstruction distortion at Eve under a minimum-leakage constraint. The resulting bit-flip probabilities serve as the secrecy protection requirements against Eve.
        
        \item We formulate a joint power and AN allocation problem that incorporates the learned targets for a MISO wiretap channel under a worst-case angular-separation model of Eve's channel. We show that the optimal information-to-AN power ratio for each bit admits a closed-form solution, while the bit-level power allocation is obtained by water-filling. We further develop a leakage-aware method that selects one of multiple learned target profiles according to the channel conditions.

        \item We extend the proposed design to MIMO wiretap channels by mapping groups of semantic bits to the eigenmode streams of Bob's channel according to their reliability targets. Under a diagonal approximation of the overlap between the signal subspaces of Bob and Eve, we derive the stream powers and the AN power in closed form.
    \end{itemize}

    The remainder of the paper is organized as follows. In Sec.~\ref{Sec:system_model}, we describe the system model of a digital semantic communication scenario over a MISO wiretap channel with AN. In Sec.~\ref{Sec:SF_channel}, we present the proposed wiretap SF channel and its two-stage training procedure, which learns the bit-level reliability and secrecy targets. In Sec.~\ref{Sec:Optimizaiton}, we formulate a bit-level power and AN allocation problem, derive its closed-form and water-filling solutions, and develop a selection method among multiple candidate profiles. In Sec.~\ref{Sec:MIMO}, we extend the proposed design to MIMO wiretap channels. In Sec.~\ref{Sec:Simulation}, we provide simulation results that demonstrate the effectiveness of the proposed framework. Finally, in Sec.~\ref{Sec:Conclusion}, we present our conclusions and future research directions.

    \section{System Model}\label{Sec:system_model}

      \begin{figure*}[t]
        \centering
        {\epsfig{file=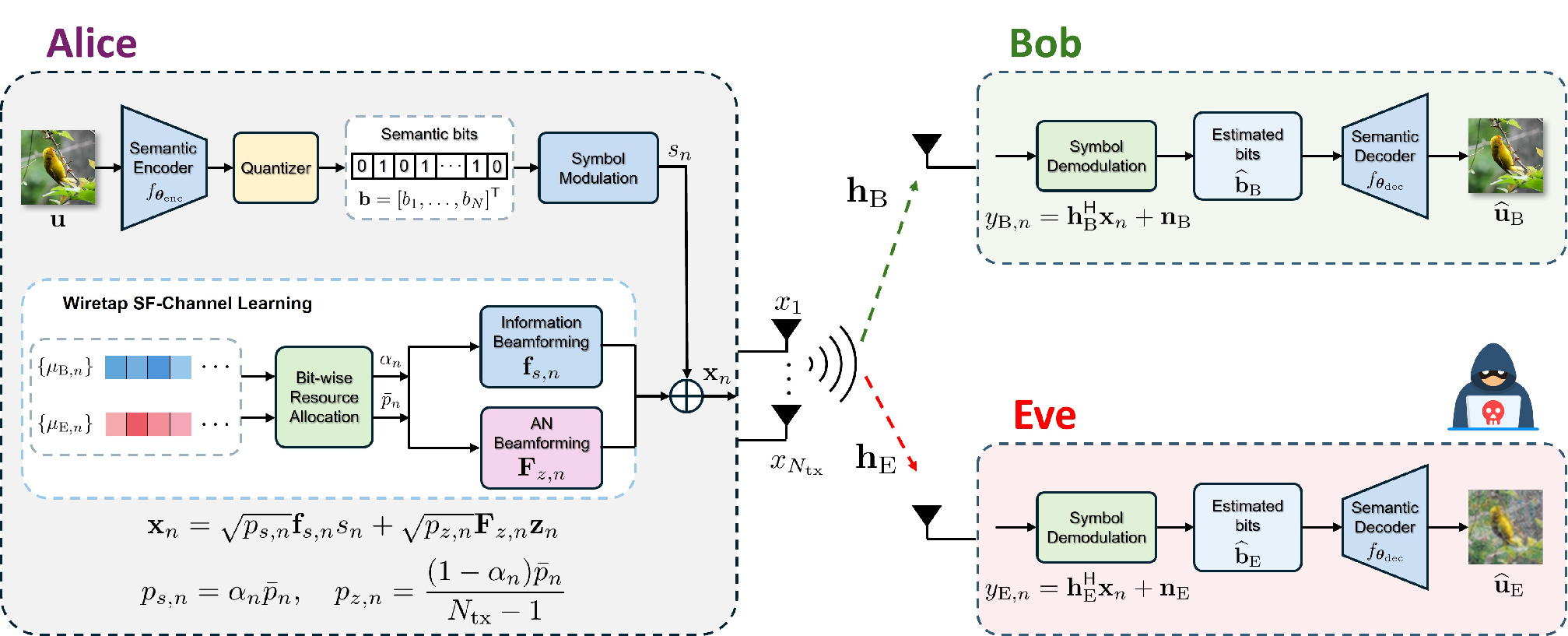,width=15cm}}\vspace{0mm}
        \caption{An illustration of a digital semantic communication system with the proposed PLS framework.}  
        \label{fig:framework}
    \end{figure*}
    
    In this work, we consider a digital semantic communication system for image transmission\footnote{In this work, we focus on an image reconstruction task for simplicity; however, the core components of the proposed framework are not limited to this setting and can be readily extended to other types of semantic sources and tasks.}, in which a transmitter (Alice) conveys task-relevant information to a legitimate receiver (Bob) in the presence of an eavesdropper (Eve). Given an image $\mathbf{u} \in \mathbb{R}^U$, Alice extracts a semantic feature from the input data using a semantic encoder, denoted by $\mathbf{b} = f_{\bm{\theta}_{\rm enc}}(\mathbf{u}) \in \{0,1\}^N$,
    where $f_{\bm{\theta}_{\rm enc}}$ denotes the semantic encoder parameterized by $\bm{\theta}_{\rm enc}$, and $\mathbf{b}=[b_1,\cdots,b_N]^\top$ is the semantic bit sequence of length $N$. In practice, the encoder produces a continuous-valued latent vector. We assume that this vector is quantized into the bit sequence $\bf{b}$ using a scalar or vector quantizer, as in \cite{oh2025digital, park2024joint}. Then, each semantic bit $b_n$ is modulated into a QAM symbol $s_n$ for transmission.\footnote{The one-to-one mapping between $b_n$ and $s_n$ corresponds to BPSK, which is assumed in our simulations. For higher-order $M$-QAM, the $\log_2 M$ bits carried by each symbol share a common power ratio and power budget, so the design is also applicable beyond BPSK, as discussed in Sec.~\ref{Sec:MIMO}.}

    We assume a MISO wiretap channel for both the Alice-to-Bob and Alice-to-Eve links. Alice is equipped with $N_{\rm tx}$ antennas, while Bob and Eve are each equipped with a single antenna. Let $\mathbf{h}_{\rm B},\mathbf{h}_{\rm E}\in\mathbb{C}^{N_{\rm tx}}$ denote the channel vectors from Alice to Bob and Eve, respectively. Without loss of generality, the channel vectors are normalized as $||\mathbf{h}_{\rm B}||^2=||\mathbf{h}_{\rm E}||^2=1$.
    We do not assume that Eve's CSI is known at Alice, which is impractical in most scenarios. Instead, we assume that Alice only knows that the angular separation between Bob and Eve is no smaller than $\theta$, which can be considered as a milder assumption on the CSI availability. Accordingly, the channel vectors satisfy
    \begin{align}
        |\mathbf{h}_{\rm B}^{\sf H}\mathbf{h}_{\rm E}| \leq \cos\theta.
    \end{align}

    To exploit the spatial degrees of freedom for security, Alice injects AN into the null space of Bob's channel\cite{goel2008guaranteeing, zhou2010secure, zhou2016physical}. More specifically, the transmitted signal for the $n$-th symbol is
    \begin{align}\label{eq:tx_signal}
        \mathbf{x}_n = \sqrt{p_{s,n}}\,\mathbf{f}_{s,n}s_n + \sqrt{p_{z,n}}\,\mathbf{F}_{z,n}\mathbf{z}_n,
    \end{align}
    where the information beamforming vector is matched to Bob's channel as $\mathbf{f}_{s,n} = \mathbf{h}_{\rm B}$,
    $[\mathbf{f}_{s,n}, \mathbf{F}_{z,n}]$ is a unitary matrix so that $\mathbf{F}_{z,n}\in\mathbb{C}^{N_{\rm tx}\times(N_{\rm tx}-1)}$ spans the null space of $\mathbf{h}_{\rm B}$, and $\mathbf{z}_n\sim\mathcal{CN}(\mathbf{0},\mathbf{I}_{N_{\rm tx}-1})$ is the AN vector. The information power and the AN power allocated to the $n$-th symbol are given by
    \begin{align}\label{eq:power_split}
        p_{s,n} = \alpha_n \bar{p}_n, \quad
        p_{z,n} = \frac{1-\alpha_n}{N_{\rm tx}-1}\bar{p}_n,
    \end{align}
    where $\alpha_n\in[0,1]$ is the information-to-AN power ratio and $\bar{p}_n$ is the bit-wise power budget assigned to the $n$-th symbol.

    The received signals at Bob and Eve are given by
    \begin{align}\label{eq:rx_signal}
        y_{{\rm B},n} = \mathbf{h}_{\rm B}^{\sf H}\mathbf{x}_n + z_{{\rm B},n}, \quad
        y_{{\rm E},n} = \mathbf{h}_{\rm E}^{\sf H}\mathbf{x}_n + z_{{\rm E},n},
    \end{align}
    where $z_{{\rm B},n}\sim\mathcal{CN}(0,\sigma_{\rm B}^2)$ and $z_{{\rm E},n}\sim\mathcal{CN}(0,\sigma_{\rm E}^2)$ denote the AWGN at Bob and Eve, respectively. Since the AN lies in the null space of $\bf{h}_{\rm B}$, it does not interfere with Bob's received signal, whereas it degrades Eve's received signal.

    At Bob, the received signals $\{y_{{\rm B},n}\}$ are equalized and demodulated to obtain an estimate $\hat{\mathbf{b}}_{\rm B}$ of the transmitted semantic bit sequence $\mathbf{b}$. The estimated bits are then fed to the semantic decoder $f_{\bm{\theta}_{\rm dec}}$, parameterized by $\bm{\theta}_{\rm dec}$, yielding the reconstructed image $\hat{\mathbf{u}}_{\rm B}=f_{\bm{\theta}_{\rm dec}}(\hat{\mathbf{b}}_{\rm B})$. We assume that Eve has full knowledge of the decoder architecture and its trained parameters $\bm{\theta}_{\rm dec}$ and uses the same decoder as Bob. Accordingly, Eve's reconstruction is modeled as $\hat{\mathbf{u}}_{\rm E}=f_{\bm{\theta}_{\rm dec}}(\hat{\mathbf{b}}_{\rm E})$, where $\hat{\mathbf{b}}_{\rm E}$ denotes Eve's estimate of $\mathbf{b}$ obtained from its received signals $\{y_{{\rm E},n}\}$.


    \section{Learning Bit-Level Reliability and Secrecy Targets of Wiretap Semantic Feature Channels}\label{Sec:SF_channel}
    In this section, we present a two-stage training strategy to learn the bit-level reliability and secrecy targets of the wiretap SF channel, parameterized by $\{\mu_{{\rm B},n},\mu_{{\rm E},n}\}$. Following the joint encoder-channel-decoder learning perspective of \cite{oh2026semantic}, we first train the encoder, decoder, and Bob's SF channel to learn the reliability requirements at Bob. Subsequently, we train Eve's SF channel with the encoder and decoder fixed to obtain the secrecy protection requirements against Eve.

    \subsection{Bit-Level Modeling for the Wiretap SF Channel}
    As in Sec.~\ref{Sec:system_model}, Alice extracts a semantic bit sequence $\mathbf{b}$ from the source image $\mathbf{u}$. Each bit $b_n$ subsequently passes through the physical wiretap channel, undergoing channel encoding, transmission, and decoding at both Bob and Eve. This end-to-end process can be equivalently represented by two independent bit-level transfer functions,
    \begin{align}
        \hat{b}_{{\rm B},n}=T_{{\rm B},n}(b_n)\in\{0,1\},\;
        \hat{b}_{{\rm E},n}=T_{{\rm E},n}(b_n)\in\{0,1\}, \;\forall n.
    \end{align}
    The characteristics of $T_{{\rm B},n}$ and $T_{{\rm E},n}$ depend on the specific communication configuration, which is unknown or not specified at the time of the training. We therefore abstract both transfer functions as independent binary symmetric channels (BSCs) with bit-error probabilities:
    \begin{align}
        \text{Pr}(\hat{b}_{{\rm B},n}\neq b_n)=\mu_{{\rm B},n}, \;
        \text{Pr}(\hat{b}_{{\rm E},n}\neq b_n)=\mu_{{\rm E},n}, \; \forall n,
    \end{align}
    as done in \cite{oh2025digital, erdemir2022privacy, NECST}. Bob then reconstructs the image as $\hat{\bf u}_{{\rm B}}=f_{\bm{\theta}_{\rm dec}}(\hat{\bf b}_{\rm B})$. As mentioned in Sec.~\ref{Sec:system_model}, Eve is assumed to possess the exact trained weights $\bm{\theta}_{\rm dec}$ of this decoder. Accordingly, we model Eve's reconstruction as the result of applying this identical, fully trained decoder to Eve's received bits, $\hat{\bf{u}}_{\rm E}=f_{\bm{\theta}_{\rm dec}}(\hat{{\bf b}}_{\rm E})$. This detail is central to the two-stage training procedure described later.

    \subsection{Motivation: Learning Secrecy Protection Requirement against Eve}\label{Sec:3_B}

    The SF channel framework in \cite{oh2026semantic} learns bit-level reliability requirements by jointly optimizing the semantic encoder, SF channel, and decoder. In the wiretap setting, we additionally seek to learn bit-level secrecy requirements against Eve. Since physical-layer security may not completely prevent information leakage to Eve, we consider a prescribed level of residual leakage during training. Specifically, we impose a positive lower bound $I_{\mathrm{E,min}}$ on the sum of Eve's BSC capacities as a leakage surrogate, rather than a lower bound on the actual mutual information.

    Under this constraint, we seek to distribute leakage across semantic bits according to their influence on Eve's reconstruction. Since Eve uses Bob's fixed decoder $f_{\bm{\theta}{\rm dec}}$, its reconstruction sensitivity varies across bits. Consequently, a less influential bit may contribute little to Eve's reconstruction even when $\mu_{{\rm E},n}<\mu_{{\rm B},n}$, as illustrated in Sec.~\ref{Sec:Sim_crossover}. We therefore maximize Eve's reconstruction distortion by optimizing ${\mu_{{\rm E},n}}$ subject to the capacity-based leakage constraint.

    \subsection{Two-Stage Training of the Wiretap SF Channel}\label{Sec:3_C}

      \begin{figure}[t]
        \centering
        {\epsfig{file=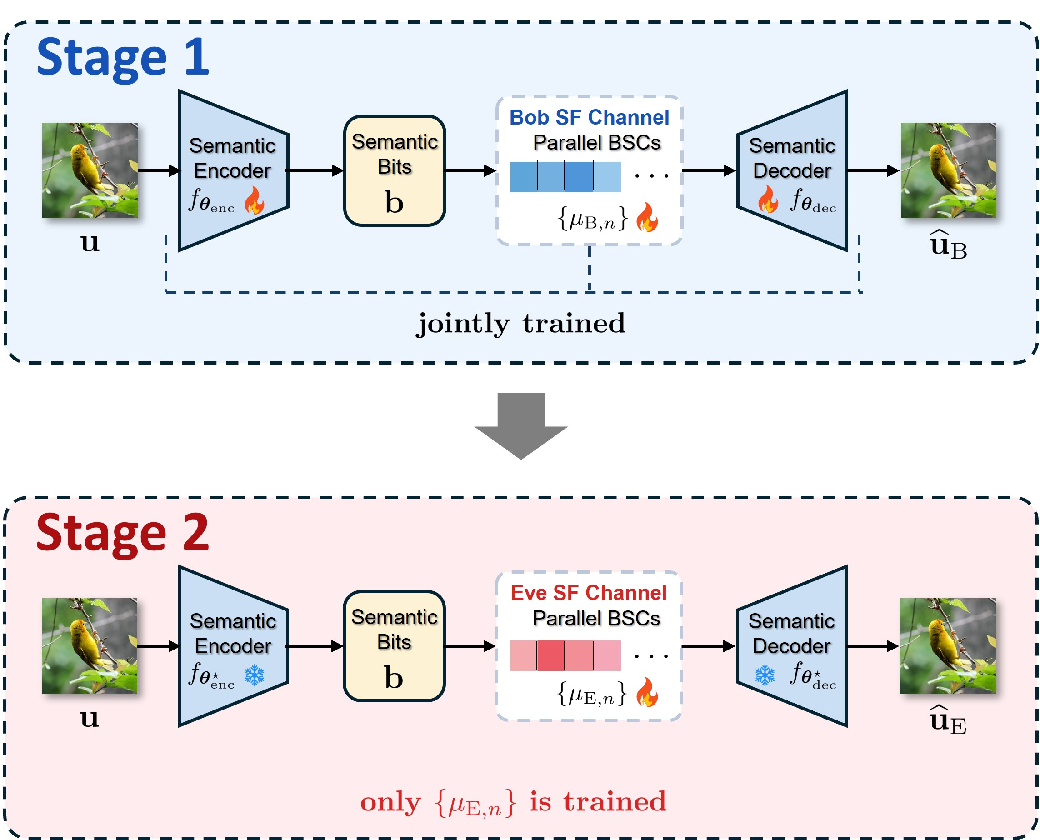,width=\columnwidth}}\vspace{0mm}
        \caption{Two-stage training of the wiretap SF channel: joint training of the encoder, decoder, and Bob's SF channel (stage 1), followed by training of Eve's SF channel with the encoder and decoder fixed (stage 2).}  
        \label{fig:two_stage_training}
    \end{figure}

    Based on the motivation discussed above, we introduce a two-stage training strategy that extends the training process in \cite{oh2026semantic} to the wiretap SF setting. In \textbf{stage 1}, the semantic encoder, decoder, and Bob's bit-flip probabilities are jointly optimized to minimize Bob's reconstruction distortion subject to an \textit{upper} bound on the information conveyed to Bob. Following the mutual information upper bound established in \cite{oh2026semantic}, the optimization problem is formulated as
    \begin{subequations}
    \begin{align}
        \min_{\bm{\theta}_{\rm enc},\bm{\theta}_{\rm dec},\{\mu_{{\rm B},n}\}} ~&
        \mathbb{E}[\|\bf{u}-\hat{\bf u}_{\rm B} \|^2] \\
        \text{s.t.}\quad\quad~&
        \sum_{n=1}^N (1-H_2(\mu_{{\rm B},n}))\leq I_{\rm B, max},\\
        &        0\leq \mu_{{\rm B},n}\leq 0.5, \; \forall n,
    \end{align}
    \end{subequations}
    where $H_2(u)\triangleq -u\log_2 u-(1-u)\log_2(1-u)$, $\hat{\bf u}_{\rm B}=f_{\bm{\theta}_{\rm dec}}(\hat{\bf b}_{\rm B})$, and $\hat{\bf b}_{\rm B}$ is obtained by passing $\bf{b}=f_{\bm{\theta}_{\rm enc}}(\bf{u})$ through the parallel BSCs whose bit-flip probabilities are $\{\mu_{{\rm B},n}\}$. The budget $I_{\rm B, max}$ prevents the trivial solution $\mu_{{\rm B},n}=0,\;\forall n$, thereby constraining the total information conveyed to Bob and encouraging the encoder and decoder to prioritize bits that are more relevant for reconstruction. This problem can be solved following the same procedure as in the single-channel setting of \cite{oh2026semantic}, yielding the optimized parameters $\bm{\theta}_{\rm enc}^{\star}$, $\bm{\theta}_{\rm dec}^{\star}$, and $\{\mu_{{\rm B},n}^{\star}\}$. 
    
    At \textbf{stage 2}, with fixed $\bm{\theta}_{\rm enc}^{\star}$ and $\bm{\theta}_{\rm dec}^{\star}$, we train only $\{\mu_{{\rm E},n}\}$. Eve's reconstruction is $\hat{\bf u}_{\rm E}=f_{\bm{\theta}_{\rm dec}^{\star}}(\hat{\bf b}_{\rm E})$, where $\hat{\bf b}_{\rm E}$ is obtained by passing $\bf{b}=f_{\bm{\theta}_{\rm enc}^{\star}}(\bf{u})$ through the parallel BSCs with $\{\mu_{{\rm E},n}\}$. Rather than minimizing the reconstruction distortion as in \textbf{stage 1}, we now maximize Eve's reconstruction distortion, subject to a \textit{lower} bound on the sum of the capacities of Eve's BSCs, which models the residual leakage:
    \begin{subequations}
    \begin{align}
        \max_{\{\mu_{{\rm E},n}\}} ~&
        \mathbb{E}[\|\bf{u}-\hat{\bf u}_{\rm E} \|^2] \\
        \text{s.t.}~~~&
        \sum_{n=1}^N (1-H_2(\mu_{{\rm E},n}))\geq I_{\rm E, min},\\
        &        0\leq \mu_{{\rm E},n}\leq 0.5, \; \forall n,
    \end{align}
    \end{subequations}
    where $I_{\rm E, min}$ specifies the prescribed minimum value of the capacity-based leakage surrogate during training. Since $1-H_2(\mu_{{\rm E},n})$ is the capacity of a BSC, this constraint does not establish a lower bound on the actual mutual information. Only $\{\mu_{{\rm E},n}\}$ are optimized in this stage; the encoder and decoder remain fixed at their optimized parameters in \textbf{stage 1}.

    As a result, our two-stage training yields the learned-target profile $\{(\mu_{{\rm B},n}^{\star},\mu_{{\rm E},n}^{\star})\}$, which characterizes the bit-level reliability requirements at Bob as well as the secrecy protection requirements against Eve. These targets are then used as the explicit input for the physical-layer resource allocation design in Sec.~\ref{Sec:Optimizaiton}.

    \section{Bit-Level Power and AN Allocation}\label{Sec:Optimizaiton}
    In this section, we develop a bit-level power and AN allocation scheme for secure semantic communication, using the reliability and secrecy targets $\{(\mu_{{\rm B},n},\mu_{{\rm E},n})\}$ learned through the training process in Sec.~\ref{Sec:SF_channel}. Specifically, we jointly optimize the power-splitting factors $\{\alpha_n\}$ and the bit-wise power allocations $\{\bar{p}_n\}$ introduced in Sec.~\ref{Sec:system_model}. The objective is to improve Bob's reception according to his learned reliability targets while enforcing the constraints on the secrecy targets against Eve under the worst-case channel model. To solve this problem efficiently, we first establish its optimization formulation and then derive a closed-form solution for $\{\alpha_n\}$ and a water-filling procedure for $\{\bar{p}_n\}$ based on a tractable objective surrogate.

    \subsection{Problem Formulation}
    Under the transmission scheme described in Sec.~\ref{Sec:system_model}, we adopt a QAM-based modulation scheme in which the bit-flip probability at a receiver takes the form \cite{fitz1994bit}:
    \begin{align}\label{eq:BER_Qfunc}
        \varepsilon_n \approx C_1 Q(\sqrt{C_2 \rho_n}),
    \end{align}
    where $\rho_n$ is the effective per-bit SNR (or SINR, in the presence of interference), and $C_1$ and $C_2$ are constants determined by the modulation order. In general, \eqref{eq:BER_Qfunc} is an approximation, but it holds with equality for the BPSK and (Gray-labeled) QPSK special cases since each bit decision in these cases reduces exactly to a one-dimensional Gaussian detection problem. Writing the bit-flip probability in the Q-function form further presumes that the total disturbance, interference-plus-noise in this case, is itself Gaussian-distributed. This assumption is satisfied automatically at both receivers in our setting since the AN vector $\mathbf{z}_{n}\sim\mathcal{CN}(\mathbf{0},\mathbf{I}_{N_{\rm tx}-1})$ is Gaussian by construction and any linear combination of jointly Gaussian random variables is itself Gaussian.

    Specializing \eqref{eq:BER_Qfunc} to Bob gives the following bit-flip probability for the $n$-th semantic bit
    \begin{align}\label{eq:BER_Bob}
        \varepsilon_{{\rm B},n}(\alpha_n,\bar{p}_n)&\approx
        C_1 Q\bigg(\sqrt{C_2 \rho_{{\rm B},n}(\alpha_n,\bar{p}_n)}\bigg),  
    \end{align}
    where $\rho_{{\rm B},n}(\alpha_n,\bar{p}_n) = {\alpha_n \bar{p}_n}/{\sigma_{\rm B}^2}$ is the effective SNR at Bob for the $n$-th bit obtained from \eqref{eq:rx_signal} under the matched beamforming $\mathbf{f}_{s,n}=\mathbf{h}_{\rm B}$. Since $||\mathbf{h}_{\rm B}^{\sf H}\mathbf{F}_{z,n}||^2=0$, the AN contributes no interference at Bob, so \eqref{eq:BER_Bob} depends only on the power allocated to the information signal, $\alpha_n\bar{p}_n$.

    At Eve, by contrast, the AN generally does not lie in the null space of $\mathbf{h}_{\rm E}$, and hence acts as an effective interference term. Since Alice has no access to Eve's CSI, we characterize security through the worst-case (i.e., most favorable to Eve) SINR over (i) all channel realizations consistent with the assumed minimum angular separation $|\mathbf{h}_{\rm B}^{\sf H}\mathbf{h}_{\rm E}| \leq \cos \theta$, and (ii) all noise levels consistent with an assumed floor $\sigma_{\rm E}^2 \geq \sigma_{\rm min}^2\triangleq 0$ on Eve's receiver noise variance. This gives the following upper bound on Eve's SINR:
    \begin{align}
        \rho_{{\rm E},n}&=\frac{p_{s,n}|\mathbf{h}_{\rm E}^{\sf H}\mathbf{f}_{s,n}|^2}{p_{z,n}\|\mathbf{h}_{\rm E}^{\sf H}\mathbf{F}_{z,n}\|^2+\sigma_{\rm E}^2} \nonumber \\
        &=\frac{p_{s,n}|\mathbf{h}_{\rm E}^{\sf H}\mathbf{f}_{s,n}|^2}{p_{z,n}(\|\mathbf{h}_{\rm E}\|^2-|\mathbf{h}_{\rm E}^{\sf H}\mathbf{f}_{s,n}|^2)+\sigma_{\rm E}^2} \nonumber \\
        &\leq \frac{\alpha_n \bar{p}_n \cos^2\theta}{\frac{1}{N_{\rm tx}-1}(1-\alpha_n)\bar{p}_n(1-\cos^2\theta)+\sigma_{\rm min}^2} \nonumber \\
        &= \frac{(N_{\rm tx}-1)\cos^2\theta}{1-\cos^2\theta}\cdot \frac{\alpha_n}{1-\alpha_n} \triangleq \rho_{{\rm E},n}^{\rm (ub)}(\alpha_n),
    \end{align}
    where the corresponding worst-case bit-flip probability is given by
    \begin{align}\label{eq:BER_Eve}
        \varepsilon_{{\rm E},n}^{\rm (lb)}(\alpha_n)=C_1Q\bigg(\sqrt{C_2\rho_{{\rm E},n}^{\rm (ub)}(\alpha_n)} \bigg).
    \end{align}

    Notably, \eqref{eq:BER_Eve} depends only on $\alpha_n$ and not on $\bar{p}_n$. Specifically, increasing $\bar{p}_n$ scales both the information and AN powers proportionally through \eqref{eq:power_split}, leaving their ratio, and hence Eve's worst-case SINR, unchanged. Consequently, $\alpha_n$ alone governs the reliability-secrecy trade-off for the $n$-th bit, while $\bar{p}_n$ governs how the total power budget is distributed across bits.

    Rather than minimizing the sum of the achieved bit-flip probabilities $\varepsilon_{{\rm B},n}$ directly, we measure the reliability of each bit by the cross-entropy between the Bernoulli distributions of the achieved bit-flip probability and its learned target,
    \begin{align}\label{eq:cross_entropy}
        H(\varepsilon_{{\rm B},n}, \mu_{{\rm B},n}) =&
        -\varepsilon_{{\rm B},n}\log_2 \mu_{{\rm B},n} \nonumber \\
        &-(1-\varepsilon_{{\rm B},n})\log_2(1-\mu_{{\rm B},n}),
    \end{align}
    and minimize the sum of these cross-entropies subject to Eve's worst-case bit-flip probability meeting its target secrecy level exactly and the total power budget $P_{\rm tot}$:
    \begin{subequations}
    \begin{align}\label{eq:P1}
        {\textbf{(P1)}} \quad \min_{\{\alpha_n,\bar{p}_n\}} ~&
        \sum_{n=1}^N H(\varepsilon_{{\rm B},n}, \mu_{{\rm B},n}) \\
        \text{s.t.}~~~&
        \varepsilon_{{\rm E},n}^{\rm (lb)}(\alpha_n)=\mu_{{\rm E},n},\;\forall n, \label{eq:P1_C1}\\
        &        \sum_{n=1}^N \bar{p}_n =P_{\rm tot},~~\bar{p}_n \geq 0, \;\forall n,\label{eq:P1_C2}\\
         &        \alpha_n \in [0,1],\;\forall n. \label{eq:P1_C3} 
    \end{align}
    \end{subequations}

    Expanding \eqref{eq:cross_entropy} gives $H(\varepsilon_{{\rm B},n}, \mu_{{\rm B},n}) =\omega_{n}\varepsilon_{{\rm B},n}-\log_2 (1-\mu_{{\rm B},n})$, where the second term acts as a constant during optimization since $\mu_{{\rm B},n}$ is a fixed, pre-learned target, and
    \begin{align}
        \omega_{n}\triangleq\log_2\frac{1-\mu_{{\rm B},n}}{\mu_{{\rm B},n}}.
    \end{align}
    Substituting this into \textbf{(P1)} and discarding the constant term shows that \textbf{(P1)} is equivalent to minimizing the weighted bit-flip probabilities $\sum_{n=1}^N \omega_{n} \varepsilon_{{\rm B},n}(\alpha_n,\bar{p}_n)$ subject to the same constraints, where the weight $\omega_n$ serves to prioritize bits according to the reliability targets at Bob, arising directly from the cross-entropy objective in \textbf{(P1)} rather than being chosen heuristically. Since $\omega_n$ is monotonically decreasing in $\mu_{{\rm B},n}$, bits with more stringent requirements at Bob are assigned correspondingly larger weights, so that the resulting optimization preferentially allocates power to protect those bits. 

    To make the optimization problem more tractable, we adopt the Chernoff bound $Q(x)\leq \frac{1}{2}e^{-x^2/2}$ on Bob's bit-flip probability \eqref{eq:BER_Bob}. Then, the weighted bit-flip probability $\omega_n \varepsilon_{{\rm B},n}(\alpha_n,\bar{p}_n)$ admits the tractable surrogate
    \begin{align}
        \omega_n \varepsilon_{{\rm B},n}(\alpha_n,\bar{p}_n)\leq
        \frac{C_1}{2}\omega_{n}\exp(-A\alpha_n\bar{p}_n),
    \end{align}
    where $A\triangleq C_2/(2\sigma_{\rm B}^2)$. It preserves the monotonic dependence on $\alpha_n\bar{p}_n$ while being amenable to closed-form optimization. Combining this surrogate with the weighted reduction above gives the following tractable surrogate problem for \textbf{(P1)}, used to compute the resource allocation:
    \begin{subequations}
    \begin{align}\label{eq:P1_prime}
        {\textbf{(P1)$^{\prime}$}} \quad \min_{\{\alpha_n,\bar{p}_n\}} ~&
        \sum_{n=1}^N \omega_n \exp(-A\alpha_n\bar{p}_n) \\
        \text{s.t.}~~~&
        \eqref{eq:P1_C1}\text{--}\eqref{eq:P1_C3}.
    \end{align}
    \end{subequations}

    \subsection{Solution to the Bit-Level Security Problem}

    Since \eqref{eq:P1_C1} fixes $\varepsilon_{{\rm E},n}^{\rm (lb)}(\alpha_n)$ exactly at $\mu_{{\rm E},n}$ for each $n$, $\alpha_n$ is determined directly by this equality, independently of $\{\bar{p}_n\}$. Solving $\varepsilon_{{\rm E},n}^{\rm (lb)}(\alpha_n)=\mu_{{\rm E},n}$ for $\alpha_n$ gives the closed-form solution
    \begin{align}\label{eq:alpha_solution}
        \alpha_n^{\star}=\left(1+\frac{C_2 (N_{\rm tx}-1)\cos^2 \theta}{(1-\cos^2\theta)\big(Q^{-1}(\mu_{{\rm E},n}/C_1)\big)^2} \right)^{-1}.
    \end{align}
    Since $Q^{-1}(\cdot)$ is strictly decreasing, a more stringent secrecy requirement (larger $\mu_{{\rm E},n}$, demanding a higher guaranteed bit-flip probability at Eve) yields a smaller $Q^{-1}(\mu_{{\rm E},n}/C_1)$ and hence a smaller $\alpha_n^{\star}$: bits that demand stronger secrecy are allocated proportionally more power to the AN component, which is precisely the intended behavior of the bit-level security design. Moreover, $\alpha_n^{\star}$ in \eqref{eq:alpha_solution} naturally satisfies \eqref{eq:P1_C3}.

    With $\{\alpha_n^{\star}\}$ fixed via \eqref{eq:alpha_solution}, the original problem {\textbf{(P1)}} reduces to allocating the total power budget $P_{\rm tot}$ across bits, which is formulated as below:
    \begin{subequations}
    \begin{align}\label{eq:P2}
        {\textbf{(P2)}} \quad \min_{\{\bar{p}_n\}} ~&
        \sum_{n=1}^N \omega_n \exp(-A\alpha_n^{\star}\bar{p}_n) \\
        \text{s.t.}~~&
        \sum_{n=1}^N \bar{p}_n =P_{\rm tot},~~\bar{p}_n \geq 0, \;\forall n. \label{eq:P2_C1}
    \end{align}
    \end{subequations} 
    Since $\exp(-A\alpha_n^{\star}\bar{p}_n)$ is convex in $\bar{p}_n$ for each $n$, the objective in \textbf{(P2)} is a nonnegatively weighted sum of convex functions and is therefore convex. Meanwhile, the power budget and non-negativity constraints define a convex feasible set. Therefore, Problem \textbf{(P2)} is a convex optimization problem whose optimal solution can be characterized through the KKT conditions, which take the familiar form of a water-filling problem. This is stated in the following proposition:
    \vspace{1mm}
    \begin{prop}\label{prop:water_filling}
        The optimal solution to \textbf{(P2)} is given by
        \begin{align}\label{eq:water_filling}
            \bar{p}_n^{\star}=\left[\frac{1}{A\alpha_n^{\star}}\ln\left(\frac{A\alpha_n^{\star}\omega_n}{\nu}\right) \right]^{+},\;[x]^{+}\triangleq\max(x,0),
        \end{align}
        where $\nu>0$ is the Lagrange multiplier associated with the power budget constraint, chosen such that $\sum_{n=1}^N \bar{p}_n^{\star}=P_{\rm tot}$.
    \end{prop}

    \begin{IEEEproof}
        See Appendix~\ref{Apdx:Prop1}.
    \end{IEEEproof}
    \vspace{1mm}
    
    Since $\bar{p}_n(\nu)$ in \eqref{eq:water_filling} is monotonically decreasing in $\nu$, the optimal value of $\nu$ can be efficiently determined using a bisection search with proper starting points $\nu^{\rm (lb)}$ and $\nu^{\rm (ub)}$. For example, $\nu^{\rm (ub)}=\max_n A \alpha_n^{\star} \omega_n$ and $\nu^{\rm (lb)}=A\alpha_1^{\star} \omega_1 \exp(-2P_{\rm tot}A\alpha_1^{\star})$ give $\sum_n\bar{p}_n(\nu^{\rm (ub)})=0<P_{\rm tot}$ and $\sum_n\bar{p}_n(\nu^{\rm (lb)})>P_{\rm tot}$, respectively. The overall procedure of the proposed power and AN allocation is summarized in {\bf Algorithm~\ref{alg:PowerAN}}.

    \begin{algorithm}[t]
        \caption{Bit-Level Power and AN Allocation}\label{alg:PowerAN}
        {\small \begin{algorithmic}[1]
            \REQUIRE $\{\mu_{{\rm B},n}\}_{n=1}^N,\, \{\mu_{{\rm E},n}\}_{n=1}^N,\, N_{\rm tx},\, \theta,\, C_1,\, C_2,\, \sigma_{\rm B}^2,\, P_{\rm tot}$, tolerance $\tau>0$
            \STATE \textbf{Compute reliability-aware weights:}
            \STATE $\omega_n \gets \log_2\left(\dfrac{1-\mu_{{\rm B},n}}{\mu_{{\rm B},n}}\right), \; \forall n$
            \STATE \textbf{Compute information-to-AN power ratios:}
            \STATE $\alpha_n^\star \gets \left(1+\dfrac{C_2(N_{\rm tx}-1)\cos^2\theta}{(1-\cos^2\theta)\big(Q^{-1}(\mu_{{\rm E},n}/C_1)\big)^2}\right)^{-1}, \; \forall n$
            \STATE $A \gets C_2/(2\sigma_{\rm B}^2)$
            \STATE \textbf{Bisection search for $\nu$:}
            \STATE Initialize $\nu^{\rm (lb)}$ and $\nu^{\rm (ub)}$
            \REPEAT
            \STATE $\nu \gets \left(\nu^{\rm (lb)}+\nu^{\rm (ub)}\right)/2$
            \STATE $\bar p_n \gets \left[\dfrac{1}{A\alpha_n^\star}\ln\left(\dfrac{A\alpha_n^\star\omega_n}{\nu}\right)\right]^+, \; \forall n$
            \IF {$\sum_{n=1}^N \bar p_n > P_{\rm tot}$}
            \STATE $\nu^{\rm (lb)} \gets \nu$
            \ELSE
            \STATE $\nu^{\rm (ub)} \gets \nu$
            \ENDIF
            \UNTIL{$\left|\sum_{n=1}^N \bar p_n - P_{\rm tot}\right| < \tau$}
            \ENSURE $\{\alpha_n^\star\}_{n=1}^N,\, \{\bar p_n\}_{n=1}^N$
        \end{algorithmic}}
    \end{algorithm}

    \subsection{Selection Among Multiple Candidate Profiles}\label{Sec:ProfileSelection}
    The two-stage training in Sec.~\ref{Sec:3_C} produces a single learned-target profile $\{(\mu_{{\rm B},n},\mu_{{\rm E},n})\}$ for a given choice of the information budgets $I_{\rm B,max}$ and $I_{\rm E,min}$. In practice, however, the appropriate values of $I_{\rm B,max}$ and $I_{\rm E,min}$ are not known at training time. It depends on the actual channel realizations and the communication resources available at deployment, none of which the offline training pipeline has access to. We therefore train $K$ candidate target profiles, $\{(\mu_{{\rm B},n}^{(k)},\mu_{{\rm E},n}^{(k)})\}$ for $k\in[K]\triangleq\{1,\ldots,K\}$, each obtained by repeating the two-stage training in Sec.~\ref{Sec:3_C} with respect to a certain budget pair $(I_{\rm B,max}^{(k)},I_{\rm E,min}^{(k)})$. 
    This leads us to raise a new question related to the encoder-decoder deployment: 
    
    {\bf Question:} \emph{Given the channel conditions and a fixed total transmit power, which profile (and the corresponding encoder-decoder pair) should be selected, and how should its signal and AN powers be allocated?}
    
    A profile with smaller Eve's target bit-flip probabilities generally permits a larger information-signal power fraction, but also exposes more information through its modeled Eve channel. To  explicitly account for this cost, we define the normalized leakage surrogate as
    \begin{align}\label{eq:profile_leakage}
        \mathcal L_{\rm E}^{(k)}\triangleq
        \frac{1}{N}\sum_{n=1}^N\left[1-H_2\!\left(\mu_{{\rm E},n}^{(k)}\right)\right].
    \end{align}
    This quantity averages the capacities of the learned BSCs and lies in $[0,1]$. Under independent BSC errors, it upper-bounds the normalized mutual information between the input and output bit vectors, with equality for independent uniform input bits. It is used here as a channel-level leakage surrogate, not as an exact measure of semantic information leakage or Eve's reconstruction quality. Since $I_{\rm E,min}^{(k)}$ is only a lower bound used in training, the penalty is computed from the learned probabilities rather than from $I_{\rm E,min}^{(k)}$. Since binary entropy increases on $[0,1/2]$, the leakage surrogate decreases with each Eve target bit-flip probability over this interval. Thus, a smaller target incurs a larger leakage cost when the other targets are held fixed.

    We formulate the joint design using the cross-entropy in \eqref{eq:cross_entropy}, without replacing it by the exponential surrogate. For brevity, let $\varepsilon_{{\rm B},n}^{(k)}\triangleq\varepsilon_{{\rm B},n}(\alpha_n^{(k)},\bar p_n^{(k)})$, evaluated using the Q-function model in \eqref{eq:BER_Bob}. With a leakage-importance weight $\lambda\geq0$, the problem is formulated as
    \begin{subequations}\label{eq:P3}
    \begin{align}
        \textbf{(P3)}~
        \min_k \min_{\{\alpha_n^{(k)},\,\bar p_n^{(k)}\}}~&
        \sum_{n=1}^{N}H\!\left(\varepsilon_{{\rm B},n}^{(k)},\mu_{{\rm B},n}^{(k)}\right)
        +\lambda\mathcal L_{\rm E}^{(k)}\label{eq:P3_obj}\\
        \text{s.t.}\quad~~&
        \varepsilon_{{\rm E},n}^{\rm (lb)}\!\left(\alpha_n^{(k)}\right)
        =\mu_{{\rm E},n}^{(k)},~\forall n,\label{eq:P3_C1}\\
        &\sum_{n=1}^{N}\bar p_n^{(k)}=P_{\rm tot},~~\bar p_n^{(k)}\geq0,~\forall n, \label{eq:P3_C2}\\
        &0\leq\alpha_n^{(k)}\leq1,~\forall n.\label{eq:P3_C3}
        \end{align}
    \end{subequations}
    As in \textbf{(P1)}, every candidate uses the same total power $P_{\rm tot}>0$. The parameter $\lambda$ controls the trade-off between Bob's reliability and the leakage surrogate; it also accounts for their different scales.

    For a fixed $k$, the leakage penalty $\lambda\mathcal L_{\rm E}^{(k)}$ is constant with respect to the resource allocation, and the remaining problem is precisely the single-profile design \textbf{(P1)}. Applying the same cross-entropy decomposition and Chernoff bound yields \textbf{(P2)} with the targets replaced by $\{(\mu_{{\rm B},n}^{(k)},\mu_{{\rm E},n}^{(k)})\}$. Thus, \textbf{Algorithm~\ref{alg:PowerAN}} can be applied directly to each candidate, without deriving a separate water-filling solution. Denote its outputs by $\{\alpha_n^{(k),\star},\widehat p_n^{(k)}\}$; the allocation for a fixed profile is independent of $\lambda$. Throughout, we assume $0<\mu_{{\rm B},n}^{(k)}\leq1/2$ and $0<\mu_{{\rm E},n}^{(k)}/C_1<1/2$, so that $\omega_n\geq0$ and $\alpha_n^{(k),\star}\in(0,1)$.

    The profiles are ranked using the original cross-entropy, not its exponential bound. Specifically, we evaluate
    \begin{align}\label{eq:profile_evaluated_ber}
        \widehat\varepsilon_{{\rm B},n}^{(k)}
        =C_1Q\!\left(\sqrt{2A\alpha_n^{(k),\star}\widehat p_n^{(k)}}\right),
    \end{align}
    and compute
    \begin{align}\label{eq:profile_exact_cost}
        J^{(k)}=\sum_{n=1}^N H\!\left(\widehat\varepsilon_{{\rm B},n}^{(k)},\mu_{{\rm B},n}^{(k)}\right)
        +\lambda\mathcal L_{\rm E}^{(k)}.
    \end{align}
    We then select the index $\widehat k$ given by 
    \begin{align}\label{eq:profile_selected_index}
        \widehat k = \underset{k\in\{1,\ldots,K\}}{\arg\!\min}~J^{(k)}.
    \end{align}
    Since \textbf{Algorithm~\ref{alg:PowerAN}} optimizes the surrogate \textbf{(P2)}, \eqref{eq:profile_selected_index} selects the best of the computed allocations, not necessarily the global optimum of \textbf{(P3)}. \textbf{Algorithm~\ref{alg:ProfileSelection}} summarizes this outer selection procedure.


    \begin{algorithm}[t]
        \caption{Leakage-Aware Profile Selection}\label{alg:ProfileSelection}
        {\small \begin{algorithmic}[1]
            \REQUIRE $\{\mu_{{\rm B},n}^{(k)},\mu_{{\rm E},n}^{(k)}\}_{n,k}$, $N_{\rm tx}$, $\theta$, $C_1$, $C_2$, $\sigma_{\rm B}^2$, $P_{\rm tot}>0$, $\lambda\geq0$
            \FOR{$k = 1$ to $K$}
                \STATE Compute $\mathcal L_{\rm E}^{(k)}$ using \eqref{eq:profile_leakage}
                \STATE Run \textbf{Algorithm~\ref{alg:PowerAN}} for profile $k$ to obtain $\{\alpha_n^{(k),\star},\widehat p_n^{(k)}\}_{n}$
                \STATE Evaluate $\widehat\varepsilon_{{\rm B},n}^{(k)}$ using \eqref{eq:profile_evaluated_ber}, $\forall n$
                \STATE Compute $J^{(k)}$ using the exact cross-entropy cost \eqref{eq:profile_exact_cost}
            \ENDFOR
            \STATE $\widehat k\gets{\arg\!\min}_{k\in\{1,\ldots,K\}}J^{(k)}$
            \ENSURE $\widehat k$, $\{\alpha_n^{(\widehat k),\star}\}_{n=1}^N$, $\{\widehat p_n^{(\widehat k)}\}_{n=1}^N$
        \end{algorithmic}}
    \end{algorithm}

    \section{Extension to Stream-Level Design \\ for MIMO Wiretap Channels}\label{Sec:MIMO}

    In this section, we illustrate how the bit-level design in Sec.~\ref{Sec:Optimizaiton} for the MISO wiretap channel extends to a MIMO setting. Unlike the MISO case, where the worst-case secrecy bound was obtained rigorously, the MIMO extension presented here relies on a simplifying diagonal approximation of the subspace overlap between Bob's and Eve's channels. We therefore present this section as an \textit{application} of the proposed design methodology to the multi-stream scenario. 

    \subsection{MIMO Communication Scenario}

     We assume Alice is equipped with $N_{\rm tx}$ antennas, while Bob and Eve are now equipped with $N_{\rm rx}$ antennas, with $N_{\rm tx}\geq N_{\rm rx}$. Alice transmits $N_{s}=N_{\rm rx}$ parallel streams $\mathbf{s}=[s_1,\ldots,s_{N_{s}}]^{\top}$, each carrying a group of semantic bits, together with an AN vector. Let $\mathbf{H}_{\rm B},\mathbf{H}_{\rm E}\in\mathbb{C}^{N_{\rm rx}\times N_{\rm tx}}$ denote the Alice-to-Bob and Alice-to-Eve MIMO channels, respectively, normalized such that $\mathbb{E}[\|\mathbf{H}_{\rm B}\|_{F}^2]=\mathbb{E}[\|\mathbf{H}_{\rm E}\|_{F}^2]=N_{\rm rx}$, which is consistent with the unit-norm channel vectors of the MISO case for $N_{\rm rx}=1$. The singular value decomposition (SVD) of $\mathbf{H}_{\rm B}$ is $\mathbf{H}_{\rm B}=\mathbf{U}_{\rm B}[\bm{\Lambda}_{\rm B},\mathbf{O}]\mathbf{V}_{\rm B}^{\sf H}$,
     where $\bm{\Lambda}_{\rm B}=\text{diag}(\lambda_{{\rm B},1},\ldots, \lambda_{{\rm B},N_{\rm rx}})$ collects the singular values in descending order.

     We define the transmit beamforming matrix $\mathbf{F}=[\mathbf{F}_s,\mathbf{F}_z]$, where $\mathbf{F}_s=[\mathbf{v}_{{\rm B},1},\ldots, \mathbf{v}_{{\rm B},N_{s}}]\in \mathbb{C}^{N_{\rm tx}\times N_{s}}$ and $\mathbf{F}_z=[\mathbf{v}_{{\rm B},N_{s}+1},\ldots,\mathbf{v}_{{\rm B},N_{\rm tx}}]\in \mathbb{C}^{N_{\rm tx}\times N_{z}}$ with $N_z\triangleq N_{\rm tx}-N_s$, so that $\mathbf{F}_s$ consists of the $N_s$ dominant right singular vectors of $\mathbf{H}_{\rm B}$ while $\mathbf{F}_z$ spans the orthogonal complement of the subspace occupied by $\mathbf{F}_s$ within $\mathbf{V}_{\rm B}$. This is the direct MIMO generalization of the matched beamforming and null-space AN injection used in the MISO channel of Sec.~\ref{Sec:system_model}. The transmitted signal is 
     \begin{align} \mathbf{x}=\mathbf{F}_s\mathbf{P}^{1/2}\mathbf{s}+\sqrt{p_z}\mathbf{F}_z\mathbf{z}\in\mathbb{C}^{N_{\rm tx}},
     \end{align}
     where $\mathbf{P}=\text{diag}(p_1,\ldots,p_{N_s})$ is the per-stream information power, $\mathbf{z}\sim\mathcal{CN}(\mathbf{0},\mathbf{I}_{N_z})$ is the AN vector, and $p_z$ is the per-dimension AN power. Each stream carries one semantic bit per channel use (e.g., with BPSK), so the $N$ bits are delivered over $T\triangleq N/N_s$ channel uses within one channel coherence block. To keep the same total power $P_{\rm tot}$ over the $N$ bits as in the MISO case, the power per channel use is constrained by
     \begin{align}\label{eq:MIMO_power}
         \sum_{i=1}^{N_s}p_i+N_zp_z\leq P_{\rm c}\triangleq \frac{N_s}{N}P_{\rm tot},
     \end{align}
     so that $TP_{\rm c}=P_{\rm tot}$.

     At Bob, the received signal $\mathbf{y}_{\rm B}=\mathbf{H}_{\rm B}\mathbf{x}+\mathbf{n}_{\rm B}$ is processed with the matched receive combiner $\mathbf{W}_{\rm B}=[\mathbf{u}_{{\rm B},1},\ldots,\mathbf{u}_{{\rm B},N_s}]^{\sf H}\in\mathbb{C}^{N_s \times N_{\rm rx}}$. Since the columns of $\mathbf{F}_z$ lie in the null space of $\mathbf{H}_{\rm B}$, the AN component is exactly nulled at Bob, yielding $N_s$ decoupled streams
     \begin{align}
         \mathbf{W}_{\rm B}\mathbf{y}_{\rm B}=\bm{\Lambda}_{\rm B}\mathbf{P}^{1/2}\mathbf{s}+\mathbf{W}_{\rm B}\mathbf{n}_{\rm B},
     \end{align}
     where the $i$-th stream's SNR is $\rho_{{\rm B},i}=p_i\lambda_{{\rm B},i}^2/\sigma_{\rm B}^2$.

     \subsection{Stream-Level Mapping and Worst-Case Secrecy Analysis at Eve}
    To extend the bit-level design in Sec.~\ref{Sec:Optimizaiton} to this multi-stream setting, we group the $N$ semantic bits into $N_s$ streams according to the corresponding reliability targets at Bob. Let $\pi$ be a permutation that sorts the bits in ascending order of the reliability targets, i.e., $\mu_{{\rm B},\pi(1)}\leq\mu_{{\rm B},\pi(2)}\leq\cdots\leq\mu_{{\rm B},\pi(N)}$. The $i$-th stream carries the $T$ consecutive bits
    \begin{align}\label{eq:MIMO_groups}
        \mathcal{G}_i\triangleq\{\pi((i-1)T+1),\ldots,\pi(iT)\},\quad i\in[N_s].
    \end{align}
    Define the stream-level reliability and secrecy targets as
    \begin{align}
        \bar{\mu}_{{\rm B},i} \triangleq \min_{n\in\mathcal{G}_i} \mu_{{\rm B},n},\quad
        \bar{\mu}_{{\rm E},i} \triangleq \max_{n\in\mathcal{G}_i} \mu_{{\rm E},n},
    \end{align}
    respectively.
    By construction, we have $\bar{\mu}_{{\rm B},1} \leq \bar{\mu}_{{\rm B},2} \leq \cdots \leq \bar{\mu}_{{\rm B},N_s}$. As discussed in Sec.~\ref{Sec:Optimizaiton}, a small reliability target toward Bob indicates a bit that can be interpreted as more sensitive for image reconstruction. Based on this interpretation, the $i$-th stream is transmitted along $\mathbf{v}_{{\rm B},i}$, which corresponds to the $i$-th largest singular value $\lambda_{{\rm B},i}$, so that the most sensitive bits are transmitted over the largest channel gain. At the same time, using the largest secrecy target $\bar{\mu}_{{\rm E},i}$ in each group can be considered as a conservative approach: setting the group's bit-flip probability at Eve no lower than this target ensures that every bit's secrecy target is met. Given this mapping, the power and AN resources are allocated over the streams.
    
    
     Now, consider the received signal at Eve, given by
     \begin{align}
         \mathbf{y}_{\rm E}=\mathbf{H}_{\rm E}\mathbf{x}+\mathbf{n}_{\rm E}=\mathbf{H}_{\rm E}\mathbf{F}_s\mathbf{P}^{1/2}\mathbf{s}
         +\sqrt{p_z}\mathbf{H}_{\rm E}\mathbf{F}_z\mathbf{z}+\mathbf{n}_{\rm E}.
     \end{align}
     We assume that Eve employs a zero-forcing (ZF) receiver to separate the $N_s$ streams, which is obtained as $\mathbf{W}_{\rm E}=(\tilde{\mathbf{H}}_{\rm E}^{\sf H}\tilde{\mathbf{H}}_{\rm E})^{-1}\tilde{\mathbf{H}}_{\rm E}^{\sf H}$ with $\tilde{\mathbf{H}}_{\rm E}\triangleq \mathbf{H}_{\rm E}\mathbf{F}_s$. Then, the $i$-th row ${\bf w}_{{\rm E},i}^{\sf H}$ of $\mathbf{W}_{\rm E}$ separates the $i$-th stream. The resulting SINR of the $i$-th stream at Eve is
     \begin{align}\label{eq:MIMO_EVE_SINR}
         \rho_{{\rm E},i}&=\frac{p_i}{p_z {\bf w}_{{\rm E},i}^{\sf H} {\bf H}_{\rm E}{\bf F}_z ({\bf F}_z)^{\sf H}  {\bf H}_{\rm E}^{\sf H} {\bf w}_{{\rm E},i}  + \sigma_{\rm E}^2 \|{\bf w}_{{\rm E},i}\|^2}\\
         &\leq \frac{p_i}{p_z {\bf w}_{{\rm E},i}^{\sf H} {\bf H}_{\rm E}{\bf F}_z ({\bf F}_z)^{\sf H}  {\bf H}_{\rm E}^{\sf H} {\bf w}_{{\rm E},i}  } \triangleq \rho_{{\rm E},i}^{\rm (ub)}.
     \end{align}

    Let $\mathbf{H}_{\rm E}=\mathbf{U}_{\rm E}[\bm{\Lambda}_{\rm E},\mathbf{O}]\mathbf{V}_{\rm E}^{\sf H}$ and define the subspace-overlap matrix as
    \begin{align}
        \tilde{\bf V} \triangleq {\bf V}_{\rm E}^{\sf H}{\bf V}_{\rm B} =
        \begin{bmatrix}
        \tilde{\bf V}_{11}& \tilde{\bf V}_{12} \\
        \tilde{\bf V}_{21} & \tilde{\bf V}_{22}
        \end{bmatrix}, \; \tilde{\mathbf{V}}_{11}\triangleq [\mathbf{V}_{\rm E}]_{1:N_s}^{\sf H}[\mathbf{V}_{\rm B}]_{1:N_s},
    \end{align}
    which captures how strongly the top-$N_s$ right singular subspace of Eve's channel overlaps with Bob's channel. Then, we obtain $\tilde{\mathbf{H}}_{\rm E}^{\sf H}\tilde{\mathbf{H}}_{\rm E}=\tilde{\mathbf{V}}_{11}^{\sf H}\bm{\Lambda}_{\rm E}^2\tilde{\mathbf{V}}_{11}$, and
    \begin{align}
        &{\bf B}_{\rm E} \triangleq  {\bf W}_{\rm E}{\bf H}_{\rm E} {\bf F}_z = (\tilde{\bf V}^{\sf H}_{11}{\bm \Lambda}_{\rm E}^2 \tilde{\bf V}_{11})^{-1} \tilde{\bf V}^{\sf H}_{11}{\bm \Lambda}_{\rm E}^2 \tilde{\bf V}_{12}, \\
        &{\bf A}_{\rm E} \triangleq {\bf W}_{\rm E}{\bf H}_{\rm E} {\bf V}_{\rm B}  = [{\bf I}, {\bf B}_{\rm E}],\\
        &\|{\bf w}_{{\rm E},i}^{\sf H} {\bf H}_{\rm E}{\bf F}_z\|^2 = ({\bf B}_{\rm E}{\bf B}_{\rm E}^{\sf H})_{i,i} = ({\bf A}_{\rm E}{\bf A}_{\rm E}^{\sf H})_{i,i} - 1,\\
        & \|{\bf w}_{{\rm E},i}\|^2 = ({\bf W}_{\rm E}{\bf W}_{\rm E}^{\sf H})_{i,i}= ((\tilde{\bf V}^{\sf H}_{11}{\bm \Lambda}_{\rm E}^2 \tilde{\bf V}_{11})^{-1} )_{i,i}.
    \end{align}
    These expressions show that the singular values of Eve's channel in $\bm{\Lambda}_{\rm E}$ affect Eve's SINR only through the thermal-noise term. For a given power allocation, the AN term, and hence $\rho_{{\rm E},i}^{\rm (ub)}$, depends on the channel only through the subspace overlap, since $\mathbf{B}_{\rm E}=\tilde{\mathbf{V}}_{11}^{-1}\tilde{\mathbf{V}}_{12}$. This overlap is the direct MIMO counterpart of the scalar correlation $|\mathbf{h}_{\rm B}^{\sf H}\mathbf{h}_{\rm E}|^2$ that determined Eve's worst-case SINR in the MISO channel of Sec.~\ref{Sec:system_model}.

    \subsection{Diagonal Approximation and Closed-Form Design}
    Unlike the MISO case, where the worst-case bound follows rigorously from a scalar angular separation assumption $|\mathbf{h}_{\rm B}^{\sf H}\mathbf{h}_{\rm E}|\leq \cos \theta$, the MIMO SINR bound depends on the inverse of the subspace-overlap matrix $\tilde{\mathbf{V}}_{11}$, making the exact worst-case analysis more challenging. To retain the tractability of the design in Sec.~\ref{Sec:Optimizaiton}, we assume that every dominant right singular vector of Eve's channel is separated from every dominant right singular vector of Bob's channel by at least $\theta_{\min}$, i.e.,
    \begin{align}
        |
        \mathbf v_{{\rm E},i}^{\sf H}
        \mathbf v_{{\rm B},j}
        |
        \leq
        \cos(\theta_{\min}),
        \;
        \forall i,j.
    \end{align}
    Based on this minimum angular separation, we consider the following diagonal approximation:
    \begin{align}\label{eq:scaled_identity}
        \tilde{\mathbf{V}}_{11}=\cos(\theta_{\rm min})\cdot \mathbf{I}.
    \end{align}
    We emphasize that \eqref{eq:scaled_identity} is a simplifying approximation adopted for closed-form tractability, not a provable worst-case bound. 

    Under the diagonal approximation \eqref{eq:scaled_identity}, we obtain $\|{\bf w}_{{\rm E},i}^{\sf H} {\bf H}_{\rm E}{\bf F}_z\|^2=1/\cos^2(\theta_{\rm min})-1$ and $\|\mathbf{w}_{{\rm E},i} \|^2=1/(\lambda_{{\rm E},i}^2\cos^2(\theta_{\rm min}))$, leading to the following closed-form SINR bound:
    \begin{align}
        \rho_{{\rm E},i}^{\rm (ub)}=\frac{p_i \cos^2 (\theta_{\rm min})}{p_z(1-\cos^2(\theta_{\rm min}))}.
    \end{align}
    As in Sec.~\ref{Sec:Optimizaiton}, equating $\bar{\mu}_{{\rm E},i}=C_1Q(\sqrt{C_2\rho_{{\rm E},i}^{\rm (ub)}})$ gives
    \begin{align}
        \beta_i^{\star}\triangleq \frac{p_i}{p_z}= \frac{\big(Q^{-1}(\bar{\mu}_{{\rm E},i}/C_1)\big)^2}{C_2}\cdot \frac{1-\cos^2(\theta_{\rm min})}{\cos^2(\theta_{\rm min})}.
    \end{align}

    Since all bits in $\mathcal{G}_i$ experience the same SNR $\rho_{{\rm B},i}=p_i\lambda_{{\rm B},i}^2/\sigma_{\rm B}^2$, their bit-flip probability at Bob is $\varepsilon_{{\rm B},i}=C_1Q(\sqrt{C_2\rho_{{\rm B},i}})$. Using the stream-level target $\bar\mu_{{\rm B},i}$ for all bits in $\mathcal{G}_i$, the cross-entropy objective of Sec.~\ref{Sec:Optimizaiton} becomes $T\sum_{i=1}^{N_s}H(\varepsilon_{{\rm B},i},\bar\mu_{{\rm B},i})$. As in Sec.~\ref{Sec:Optimizaiton}, minimizing it is equivalent to minimizing $\sum_{i=1}^{N_s}\bar\omega_i\varepsilon_{{\rm B},i}$ with the stream weight $\bar\omega_i\triangleq\log_2\big((1-\bar\mu_{{\rm B},i})/\bar\mu_{{\rm B},i}\big)$. As in Sec.~\ref{Sec:Optimizaiton}, we let each stream meet its secrecy target at Eve with equality, i.e., $p_i=\beta_i^{\star}p_z$. As in the MISO case, we enforce the secrecy constraints with equality, i.e., $p_i=\beta_i^\star p_z$. Since all streams share the same AN power $p_z$, a closed-form power allocation can be derived. Applying the Chernoff bound, we reduce the design to
    \begin{subequations}
    \begin{align}
        {\textbf{(P4)}} \quad \min_{p_z\geq0} ~&
        \sum_{i=1}^{N_s} \bar\omega_i \exp\!\left(-A\lambda_{{\rm B},i}^2\beta_i^{\star} p_z\right) \label{eq:P4_obj}\\
        \text{s.t.}~~&
                \sum_{i=1}^{N_s}\beta_i^{\star} p_z +N_zp_z \leq P_{\rm c},\label{eq:P4_C1}
    \end{align}
    \end{subequations}
    where $A=C_2/(2\sigma_{\rm B}^2)$ as before. Since $\bar\omega_i\geq0$ and every term of \eqref{eq:P4_obj} is nonincreasing in $p_z$, the power constraint is active at the optimum, and \textbf{(P4)} has the optimal solution
    \begin{align}\label{eq:MIMO_power_solution}
        p_z^{\star}=\frac{P_{\rm c}}{\sum_{i=1}^{N_s} \beta_i^{\star} + N_z},\qquad
        p_i^{\star}=\beta_i^{\star}p_z^{\star},\quad i\in[N_s].
    \end{align}

    When multiple candidate profiles $\{\mu_{{\rm B},n}^{(k)},\mu_{{\rm E},n}^{(k)}\}_{k=1}^{K}$ are available, the profile selection of Sec.~\ref{Sec:ProfileSelection} applies in the same way. For each candidate $k$, the grouping $\{\mathcal{G}_i^{(k)}\}$, the stream targets $\{\bar\mu_{{\rm E},i}^{(k)}\}$, and the powers $\{p_i^{(k)}\}$ and $p_z^{(k)}$ are obtained from \eqref{eq:MIMO_groups}--\eqref{eq:MIMO_power_solution} using that profile; these quantities do not depend on the channel and can be computed offline. For each channel realization, Alice uses the known $\mathbf{H}_{\rm B}$ to evaluate Bob's bit-flip probability of the $i$-th stream as $\varepsilon_{{\rm B},i}^{(k)}=C_1Q\big(\sqrt{C_2\, p_i^{(k)}\lambda_{{\rm B},i}^2/\sigma_{\rm B}^2}\big)$ and selects the profile that minimizes the cost in \eqref{eq:profile_exact_cost}. Therefore, in the MIMO extension, the channel realization affects the power allocation only through the profile selection. The overall procedure is summarized in {\bf Algorithm~\ref{alg:MIMO}}.

     \textbf{Remark (Extension to Coded Transmission): } 
    The stream-level design above also suggests how the proposed framework can be extended to channel-coded transmission. Analogously to \eqref{eq:MIMO_groups}, the semantic bits sorted by $\pi$ are divided into $N_{\rm b}$ coding blocks of consecutive bits, $\mathcal{G}_i\triangleq \pi(\bar{k}_{i-1}+1),\ldots,\pi(\bar{k}_i))$ for $i\in[N_{\rm b}]$, where $k_i=|\mathcal{G}_i|$, $\bar{k}_i\triangleq \sum_{j=1}^{i}k_j$, $\bar{k}_0\triangleq 0$, and $\bar{k}_{N_{\rm b}}=N$. The block sizes $\{k_i\}$ are given in advance and are not optimized. The bits in $\mathcal{G}_i$ are encoded by an $(n_i,k_i)$ block code, whose codeword is transmitted over $n_i$ channel uses of the MISO channel with a common power $\bar{p}_i$ and ratio $\alpha_i$. Following \cite{kim2026unequal}, we model the block error probability of a finite block-length capacity-achieving $(n_i,k_i)$ code at the effective SINR $\rho$ by the normal approximation \cite{polyanskiy2010channel}
    \begin{align}\label{eq:cc_BLER}
        {\sf BLER}_{n_i,k_i}(\rho)=Q\left(\ln 2\sqrt{\frac{n_i}{V(\rho)}}\left(\log_2(1+\rho)-\frac{k_i}{n_i} \right) \right),
    \end{align}
    where $V(\rho)=1-(1+\rho)^{-2}$. Assuming equal and independent bit errors within a block, the bit-flip probability of each information bit can be written as $\varepsilon_{n_i,k_i}(\rho)\triangleq 1-(1-{\sf BLER}_{n_i,k_i}(\rho))^{1/k_i}$ as in \cite{kim2026unequal}. Accordingly, all bits in $\mathcal{G}_i$ experience the bit-flip probabilities $\varepsilon_{{\rm B},i}=\varepsilon_{n_i,k_i}(\rho_{{\rm B},i})$ at Bob and $\varepsilon_{{\rm E},i}^{\rm (lb)}=\varepsilon_{n_i,k_i}(\rho_{{\rm E},i}^{\rm (ub)})$ at Eve's worst case, where $\rho_{{\rm B},i}=\alpha_i\bar{p}_i/\sigma_{\rm B}^2$ and $\rho_{{\rm E},i}^{\rm (ub)}=\kappa \alpha_i/(1-\alpha_i)$ with $\kappa\triangleq (N_{\rm tx}-1)\cos^2\theta / (1-\cos^2\theta)$ as in Sec.~\ref{Sec:Optimizaiton}.
    
    As for the streams, each block is assigned the target $\bar{\mu}_{{\rm B}, i}=\min_{n\in\mathcal{G}_i}\mu_{{\rm B},n}$ and $\bar{\mu}_{{\rm E},i}=\max_{n\in\mathcal{G}_i}\mu_{{\rm E},n}$, and the design in Sec.~\ref{Sec:Optimizaiton} carries over the block level. Specifically, the cross-entropy objective becomes $\sum_{i}k_iH(\varepsilon_{{\rm B},i},\bar\mu_{{\rm B},i})$, and since $\varepsilon_{n_i,k_i}(\rho)$ is decreasing in $\rho$, Eve's constraint $\varepsilon_{{\rm E},i}^{\rm (lb)}\geq\bar\mu_{{\rm E},i}$ yields $\alpha_i^\star=\gamma_{{\rm E},i}/(\gamma_{{\rm E},i}+\kappa)$ for a given $n_i$, where $\gamma_{{\rm E},i}$ satisfies $\varepsilon_{n_i,k_i}(\gamma_{{\rm E},i})=\bar\mu_{{\rm E},i}$, analogously to \eqref{eq:alpha_solution}. The powers $\{\bar p_i\}$ are then allocated over the blocks under the total energy $\sum_{i}n_i\bar p_i\leq P_{\rm tot}$, as in \textbf{(P2)}.

    \begin{algorithm}[t]
        \caption{Stream-Level Power and AN Allocation With Profile Selection}\label{alg:MIMO}
        {\small \begin{algorithmic}[1]
            \REQUIRE $\{\mu_{{\rm B},n}^{(k)},\mu_{{\rm E},n}^{(k)}\}_{n,k}$, $N_{\rm tx}$, $N_s$, $\theta_{\rm min}$, $C_1$, $C_2$, $\sigma_{\rm B}^2$, $P_{\rm tot}$, $\lambda\geq0$, $\mathbf{H}_{\rm B}$
            \STATE $T\gets N/N_s$, $N_z\gets N_{\rm tx}-N_s$, $P_{\rm c}\gets N_sP_{\rm tot}/N$
            \STATE \textbf{Offline (channel-independent):}
            \FOR{$k = 1$ to $K$}
                \STATE Sort the bits by $\mu_{{\rm B},n}^{(k)}$ and form $\{\mathcal{G}_i^{(k)}\}_{i=1}^{N_s}$ using \eqref{eq:MIMO_groups}
                \STATE $\bar\mu_{{\rm B},i}^{(k)}\gets\min_{n\in\mathcal{G}_i^{(k)}}\mu_{{\rm B},n}^{(k)}$, $\bar\mu_{{\rm E},i}^{(k)}\gets\max_{n\in\mathcal{G}_i^{(k)}}\mu_{{\rm E},n}^{(k)}$, $\forall i$
                \STATE $\beta_i^{(k)}\gets\dfrac{\big(Q^{-1}(\bar\mu_{{\rm E},i}^{(k)}/C_1)\big)^2}{C_2}\cdot\dfrac{1-\cos^2(\theta_{\rm min})}{\cos^2(\theta_{\rm min})}$, $\forall i$
                \STATE $p_z^{(k)}\gets P_{\rm c}/\big(\sum_{i}\beta_i^{(k)}+N_z\big)$, $p_i^{(k)}\gets\beta_i^{(k)}p_z^{(k)}$, $\forall i$
                \STATE Compute $\mathcal L_{\rm E}^{(k)}$ using \eqref{eq:profile_leakage}
            \ENDFOR
            \STATE \textbf{Online (for each realization of $\mathbf{H}_{\rm B}$):}
            \STATE Compute the SVD $\mathbf{H}_{\rm B}=\mathbf{U}_{\rm B}[\bm{\Lambda}_{\rm B},\mathbf{O}]\mathbf{V}_{\rm B}^{\sf H}$; set $\mathbf{F}_s\gets[\mathbf{V}_{\rm B}]_{1:N_s}$, $\mathbf{F}_z\gets[\mathbf{V}_{\rm B}]_{N_s+1:N_{\rm tx}}$
            \FOR{$k = 1$ to $K$}
                \STATE $\varepsilon_{{\rm B},i}^{(k)}\gets C_1Q\Big(\sqrt{C_2p_i^{(k)}\lambda_{{\rm B},i}^2/\sigma_{\rm B}^2}\Big)$, $\forall i$
                \STATE $J^{(k)}\gets\sum_{i=1}^{N_s}\sum_{n\in\mathcal{G}_i^{(k)}}H\big(\varepsilon_{{\rm B},i}^{(k)},\mu_{{\rm B},n}^{(k)}\big)+\lambda\mathcal L_{\rm E}^{(k)}$
            \ENDFOR
            \STATE $\widehat k\gets\arg\min_{k}J^{(k)}$
            \STATE Transmit the bits in $\mathcal{G}_i^{(\widehat k)}$ on the $i$-th stream over $T$ channel uses with $\mathbf{x}=\mathbf{F}_s\mathbf{P}^{1/2}\mathbf{s}+\sqrt{p_z}\mathbf{F}_z\mathbf{z}$, where $\mathbf{P}={\rm diag}(p_1^{(\widehat k)},\ldots,p_{N_s}^{(\widehat k)})$ and $p_z=p_z^{(\widehat k)}$
            \ENSURE $\widehat k$, $\{p_i^{(\widehat k)}\}_{i=1}^{N_s}$, $p_z^{(\widehat k)}$
        \end{algorithmic}}
    \end{algorithm}

    \section{Simulation Results}\label{Sec:Simulation}
    In this section, we evaluate the proposed bit-level PLS framework for an image transmission task with the CIFAR-10 dataset \cite{CIFAR10}. The CIFAR-10 dataset consists of color images of size $32\times32\times3$ pixels, with 50,000 training and 10,000 test images.
\begin{table}[t]

        \renewcommand{\arraystretch}{1.1}

        {\caption{The semantic encoder and decoder architectures.}\label{table:autoencoder_architecture}}

        \setlength{\tabcolsep}{3pt}

        \scriptsize

        \centering


        \resizebox{\columnwidth}{!}{\begin{tabular}{|c|c|} \hline

            & \multicolumn{1}{c|}{Layers}   \\ \hline \hline

        \multirow{2}{*}{Encoder} & \multicolumn{1}{c|}{C(64,5,2,2), PReLU, C(128,5,2,2), PReLU,} \\

                    & \multicolumn{1}{c|}{C(128,5,1,2), PReLU, C(128,5,1,2), PReLU, C(24,5,1,2)}\\\hline

                    \multirow{2}{*}{Decoder} & \multicolumn{1}{c|}{CT(128,5,1,2,0), PReLU, CT(128,5,1,2,0), PReLU,} \\
        & \multicolumn{1}{c|}{CT(128,5,1,2,0), PReLU, CT(64,5,2,2,1), PReLU, CT(3,5,2,2,1)} \\ \hline
        

        \end{tabular}}\vspace{-2mm}


    \end{table}

    The semantic encoder-decoder pair and the wiretap SF channel are trained by the two-stage procedure in Sec.~\ref{Sec:3_C}. We employ the convolutional neural network (CNN)-based autoencoder architecture in \cite{bourtsoulatze2019deep}, detailed in Table~\ref{table:autoencoder_architecture}. In this table, ${\rm C}(c,k,s,p)$ represents a 2D convolutional layer with $c$ output channels, a $k\times k$ kernel, a stride of $s$, and padding of $p$. Similarly, ${\rm CT}(c,k,s,p,p_o)$ denotes a transposed 2D convolutional layer with output padding of $p_o$. Each encoder output is quantized with 8 bits, giving $N=12{,}288$ semantic bits. In stage 1, the encoder, the decoder, and $\{\mu_{{\rm B},n}\}$ are trained with $I_{\rm B,max}=0.9N$. In stage 2, $\{\mu_{{\rm E},n}\}$ is trained with $I_{\rm E,min}\in\{0.9N,\,0.95N,\,0.99N\}$, which yields $K=3$ candidate BSC profiles for \textbf{Algorithm~\ref{alg:ProfileSelection}}.
    For the physical layer, unless otherwise stated, Alice has $N_{\rm tx}=4$ antennas, the minimum angular separation between Bob and Eve is $\theta=60^{\circ}$, the total power is $P_{\rm tot}=N$ (i.e., $P_{\rm tot}/N=1$), and BPSK is used ($C_1=1$, $C_2=2$). The SNRs of Bob and Eve are defined as ${\rm SNR}_{\rm B}\triangleq (P_{\rm tot}/N)/(\sigma_{\rm B}^2/2)$ and ${\rm SNR}_{\rm E}\triangleq (P_{\rm tot}/N)/(\sigma_{\rm E}^2/2)$, and ${\rm SNR}_{\rm E}=20$ dB unless otherwise stated. In each transmission, $\mathbf{h}_{\rm B}$ and $\mathbf{h}_{\rm E}$ are drawn independently from a complex Gaussian distribution and normalized to unit norm, and pairs that violate the angular separation $\theta$ are redrawn.
    In the simulations, we compare the following schemes:
    \begin{itemize}
        \item \textbf{LB-Alg.~2 (learned budget with Algorithm~\ref{alg:ProfileSelection}):} This is the proposed framework. The encoder and decoder are trained in stage 1 together with Bob's bit-flip probabilities, so that the total information budget $I_{{\rm B, max}}=0.9N$ is distributed unequally over the bits. \textbf{Algorithm~\ref{alg:ProfileSelection}} runs \textbf{Algorithm~\ref{alg:PowerAN}} for each of the $K=3$ candidate profiles, selects a profile using $\lambda=1$, and returns the allocation $\{\alpha_n,\bar{p}_n\}$ of the selected profile.
        \item \textbf{LB-Unif. (learned budget with uniform allocation):} This scheme uses the same encoder and decoder as \textbf{LB-Alg.~2}, but all bits are transmitted with the same power $\bar{p}_n=P_{\rm tot}/N$ and the same ratio $\alpha_n=\alpha$, where $\alpha$ maximizes the secrecy rate under the worst-case angle. 
        \item \textbf{EB-Unif. (equal budget with uniform allocation):} This baseline does not learn bit-level targets. Its encoder and decoder are trained  over $N$ parallel BSCs with a common fixed bit-flip probability, assuming that every bit has the same information budget and the total capacity of the BSCs equals $I_{\rm B, max}=0.9 N$. 
        All bits are transmitted with the uniform allocation. 
        \item \textbf{Adv-DE (adversarial deterministic encoder):} This baseline adapts \cite{marchioro2020adversarial}, which uses a deterministic encoder. The encoder minimizes $\mathbb{E}[\|\mathbf{u}-\hat{\mathbf{u}}_{\rm B}\|^2]+\lambda_{\rm adv}\mathbb{E}[\|\bar{\mathbf{u}}-\tilde{\mathbf{u}}_{\rm E}\|^2]$, which pushes the adversarial reconstruction $\tilde{\mathbf{u}}_{\rm E}$ toward the mean training image $\bar{\mathbf{u}}$. It is trained over the MISO wiretap channel without AN at ${\rm SNR}_{\rm B}=12$ dB and ${\rm SNR}_{\rm E}=20$ dB.
         \item \textbf{Adv-SE (adversarial stochastic encoder):} This baseline adapts \cite{erdemir2022privacy}, which uses a stochastic encoder whose output bits are sampled from learned Bernoulli distributions. The encoder minimizes $\mathbb{E}[\|\mathbf{u}-\hat{\mathbf{u}}_{\rm B}\|^2]-\lambda_{\rm adv}\mathbb{E}[\|\mathbf{u}-\tilde{\mathbf{u}}_{\rm E}\|^2]$. It is trained over BSCs with bit-flip probabilities 0.1 for Bob and 0.2 for Eve, one of the settings in \cite{erdemir2022privacy}.
    \end{itemize}

    The original methods underlying \textbf{Adv-DE} and \textbf{Adv-SE} employ adversarial networks to infer source classes and sensitive attributes, respectively. To address image confidentiality, we instead use adversarial networks that reconstruct images from Eve's received bits and have the same architecture as Bob's decoder. \textbf{Adv-DE} and \textbf{Adv-SE} use the same encoder-decoder architecture and number of semantic bits as the proposed framework, and $\lambda_{\rm adv}=1$ as in \cite{marchioro2020adversarial}. At evaluation, these two baselines use uniform power and no AN as in \cite{marchioro2020adversarial, erdemir2022privacy}, and Eve uses Bob's decoder for all schemes.

    \subsection{Bit-Flip Probabilities of Bob and Eve}\label{Sec:Sim_crossover}

      \begin{figure}[t]
        \centering
        {\epsfig{file=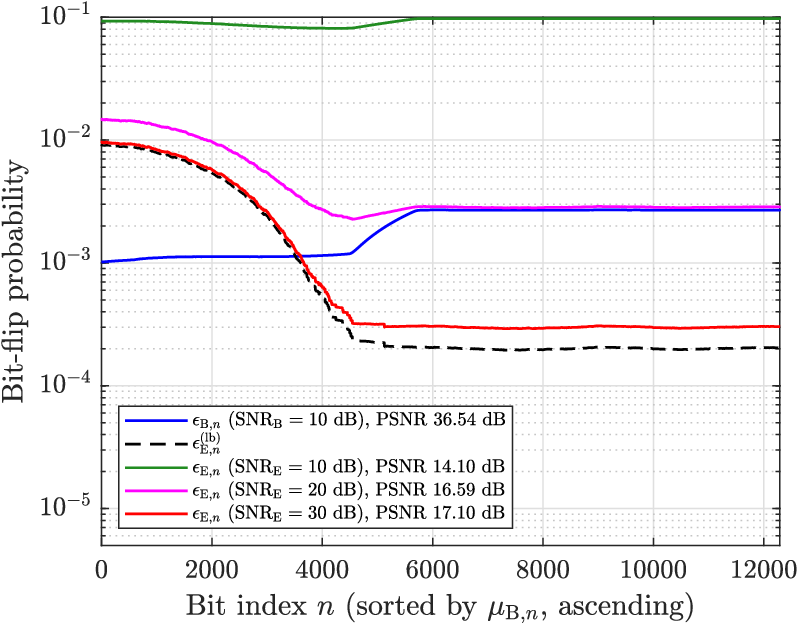,width=0.7\columnwidth}}\vspace{0mm}
        \caption{Bit-flip probabilities of Bob and Eve obtained using \textbf{Algorithm~\ref{alg:PowerAN}} with the learned wiretap SF channel on the CIFAR-10 dataset ($I_{\rm B,max}=0.9N$, $I_{\rm E,min}=0.99N$, ${\rm SNR}_{\rm B}=10$ dB, and ${\rm SNR}_{\rm E}\in\{10,20,30\}$ dB). Bits are sorted in ascending order of $\mu_{{\rm B},n}$. Each curve is a moving average over 1,176 bits, and the legend reports the corresponding PSNRs.}
        \label{fig:crossover}
    \end{figure}

    Fig.~\ref{fig:crossover} shows the bit-flip probabilities of Bob and Eve for the learned wiretap SF channel with $I_{\rm E,min}=0.99N$ at ${\rm SNR}_{\rm B}=10$ dB. Bob's bit-flip probability $\varepsilon_{{\rm B},n}$ is given by \eqref{eq:BER_Bob}, and Eve's bit-flip probability $\varepsilon_{{\rm E},n}$ is computed from the worst-case-angle SINR in Sec.~\ref{Sec:Optimizaiton} with $\sigma_{\rm min}^2$ replaced by $\sigma_{\rm E}^2$. As $\sigma_{\rm E}^2\rightarrow 0$, $\varepsilon_{{\rm E},n}$ reduces to $\varepsilon_{{\rm E},n}^{\rm (lb)}(\alpha_n^{\star})=\mu_{{\rm E},n}$ in \eqref{eq:P1_C1}. Bob's bit-flip probability is lower on the bits with small $\mu_{{\rm B},n}$, because the water-filling in \textbf{Algorithm~\ref{alg:PowerAN}} assigns more power to these bits. In contrast, Eve's lower bound is higher on these bits and decreases toward the bits with large $\mu_{{\rm B},n}$. This result demonstrates that the learned profile and the power allocation jointly prioritize reliable delivery to Bob and protection against Eve on the bits with small $\mu_{{\rm B}, n}$. 

    Fig.~\ref{fig:crossover} also compares Eve's bit-flip probabilities for ${\rm SNR}_{\rm E}\in\{10,20,30\}$ dB. At ${\rm SNR}_{\rm E}=10$ dB, Eve's bit-flip probability is nearly flat and lies above Bob's. At ${\rm SNR}_{\rm E}=20$ dB, it decreases and approaches Bob's on the bits with large $\mu_{{\rm B},n}$. At ${\rm SNR}_{\rm E}=30$ dB, it approaches the lower bound and falls below Bob's for most bits, while remaining higher on the bits with small $\mu_{{\rm B},n}$. Although Eve recovers most bits more reliably than Bob in this case, Eve's PSNR remains far below Bob's. This is because Eve's remaining errors on the bits with small $\mu_{{\rm B},n}$ dominate its reconstruction quality. This result demonstrates that steering unavoidable leakage toward the bits with large $\mu_{{\rm B},n}$ preserves a large PSNR gap between Bob and Eve.

    \subsection{PSNR Performance Versus Bob's SNR}\label{Sec:Sim_snrb}

\begin{figure}[t]
        \centering
        {\epsfig{file=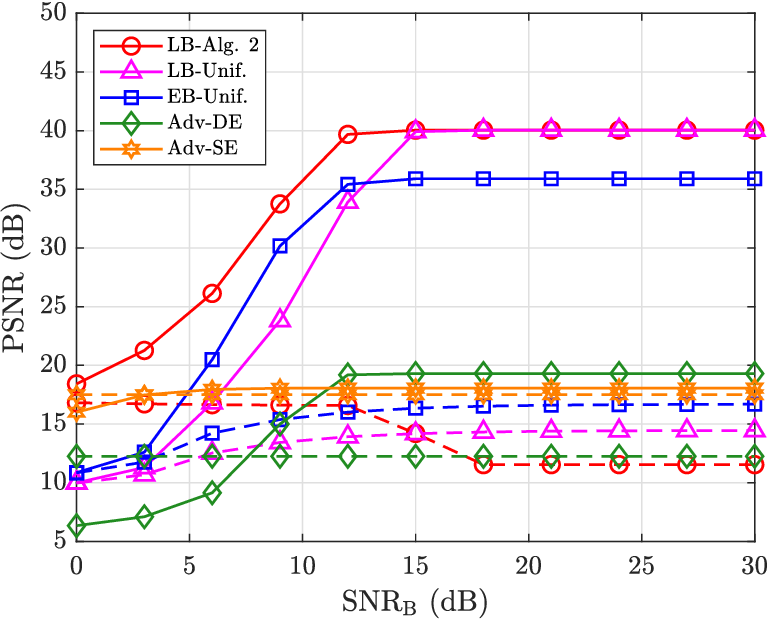,width=0.7\columnwidth}}\vspace{0mm}
        \caption{PSNRs of Bob (solid) and Eve (dashed) versus ${\rm SNR}_{\rm B}$ for ${\rm SNR}_{\rm E}=20$ dB on the CIFAR-10 dataset.}
        \label{fig:psnr_vs_snrb}
    \end{figure}

    Fig.~\ref{fig:psnr_vs_snrb} compares the PSNRs of Bob and Eve on the CIFAR-10 dataset as ${\rm SNR}_{\rm B}$ varies with ${\rm SNR}_{\rm E}=20$ dB. In the low-to-moderate ${\rm SNR}_{\rm B}$ regime, \textbf{LB-Alg.~2} achieves a higher PSNR at Bob than \textbf{LB-Unif.} and \textbf{EB-Unif.} This is because \textbf{Algorithm~\ref{alg:ProfileSelection}} allocates power according to the learned bit-level targets, favoring bits with stronger reliability requirements at Bob. In this regime, \textbf{Algorithm~\ref{alg:ProfileSelection}} selects the profile with $I_{\rm E,min}=0.99N$, which favors Bob's reliability but also yields a higher PSNR at Eve than both uniform-power schemes. As ${\rm SNR}_{\rm B}$ increases, profiles with smaller $I_{\rm E,min}$ are selected, allowing \textbf{LB-Alg.~2} to attain the same saturated PSNR at Bob as \textbf{LB-Unif.} with a lower PSNR at Eve. Although \textbf{EB-Unif.} provides a higher PSNR at Bob than \textbf{LB-Unif.} at low-to-moderate ${\rm SNR}_{\rm B}$, it saturates at a lower PSNR at Bob with a higher PSNR at Eve. This result demonstrates that the learned bit-level targets, power allocation, and profile selection jointly improve Bob's reconstruction quality and adapt the protection against Eve to Bob's channel condition.

    Fig.~\ref{fig:psnr_vs_snrb} also compares \textbf{LB-Alg.~2} with \textbf{Adv-DE} and \textbf{Adv-SE}. \textbf{LB-Alg.~2} achieves a higher PSNR at Bob over the entire ${\rm SNR}_{\rm B}$ range and a lower PSNR at Eve at high ${\rm SNR}_{\rm B}$. For both adversarial baselines, Bob's PSNR saturates at a low level because their training trades Bob's reconstruction quality for the adversarial term. This result demonstrates that the proposed framework improves Bob's reconstruction quality while also reducing Eve's reconstruction quality at high ${\rm SNR}_{\rm B}$.

      \begin{figure}[t]
        \centering
        {\epsfig{file=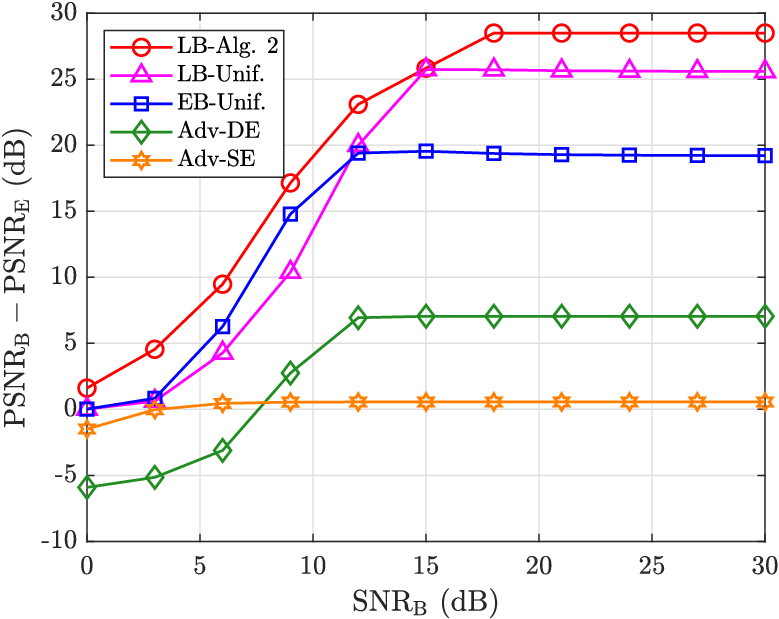,width=0.7\columnwidth}}\vspace{0mm}
        \caption{PSNR gap between Bob and Eve, ${\rm PSNR}_{\rm B}-{\rm PSNR}_{\rm E}$, versus ${\rm SNR}_{\rm B}$ for ${\rm SNR}_{\rm E}=20$ dB on the CIFAR-10 dataset.}
        \label{fig:psnr_gap_vs_snrb}
    \end{figure}

    \begin{figure}[t]
        \centering
        {\epsfig{file=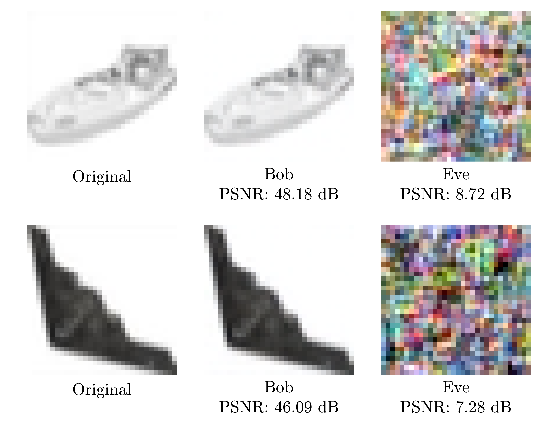,width=0.7\columnwidth}}\vspace{0mm}
        \caption{Original images and their reconstructions by Bob and Eve with \textbf{LB-Alg.~2} on the CIFAR-10 dataset (${\rm SNR}_{\rm B}=18$ dB, ${\rm SNR}_{\rm E}=20$ dB). \textbf{Algorithm~\ref{alg:ProfileSelection}} selects the profile with $I_{\rm E,min}=0.9N$. The two images have the largest PSNR gap between Bob and Eve within their classes.}
        \label{fig:recon}
    \end{figure}
    
    \begin{figure}[t]
        \centering
        {\epsfig{file=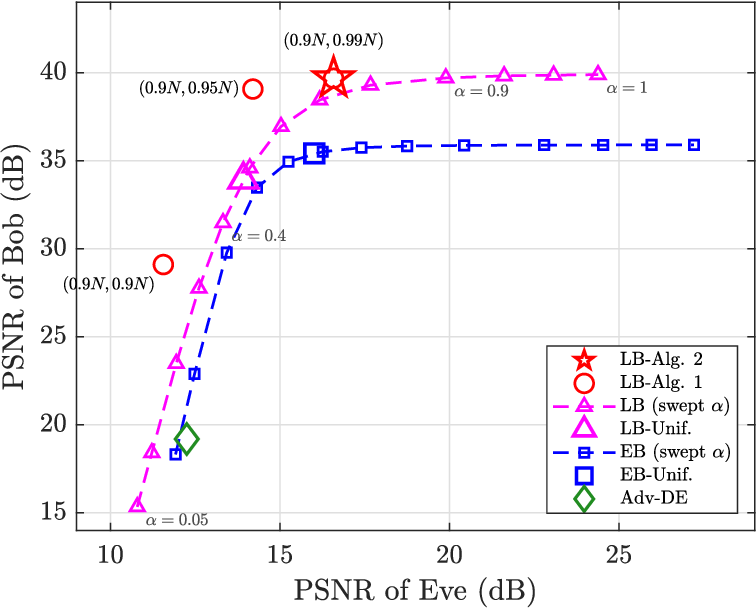,width=0.7\columnwidth}}\vspace{0mm}
        \caption{PSNR of Bob versus PSNR of Eve for ${\rm SNR}_{\rm B}=12$ dB and ${\rm SNR}_{\rm E}=20$ dB on the CIFAR-10 dataset. The upper-left corner is preferable. The dashed curves, labeled \textbf{LB (swept $\alpha$)} and \textbf{EB (swept $\alpha$)}, are obtained with the encoder-decoder pairs of \textbf{LB-Unif.} and \textbf{EB-Unif.}, respectively, using uniform power and a common ratio $\alpha$ swept from 0.05 to 1. The labels next to the \textbf{LB-Alg.~1} points indicate the profile $(I_{\rm B,max},I_{\rm E,min})$.}
        \label{fig:tradeoff}
    \end{figure}
    
    Fig.~\ref{fig:psnr_gap_vs_snrb} compares the PSNR gap between Bob and Eve over ${\rm SNR}_{\rm B}$ with ${\rm SNR}_{\rm E}=20$ dB. \textbf{LB-Alg.~2} achieves the largest gap among all evaluated schemes across the entire ${\rm SNR}_{\rm B}$ range. Compared with \textbf{LB-Unif.}, it achieves a larger gap at moderate ${\rm SNR}_{\rm B}$ by improving Bob's reconstruction quality, while its advantage at high ${\rm SNR}_{\rm B}$ comes from reducing Eve's reconstruction quality. Meanwhile, \textbf{EB-Unif.}, \textbf{Adv-DE}, and \textbf{Adv-SE} exhibit smaller saturation gaps, mainly due to their lower reconstruction quality at Bob.

    Fig.~\ref{fig:recon} shows the original images and the reconstructions obtained by \textbf{LB-Alg.~2} at ${\rm SNR}_{\rm B}=18$ dB and ${\rm SNR}_{\rm E}=20$ dB. Bob's reconstructions closely resemble the originals, whereas Eve's reconstructions contain no recognizable objects, even though Eve has a higher SNR than Bob. This result demonstrates that the proposed framework preserves Bob's reconstruction quality while degrading Eve's reconstruction.

    \subsection{PSNR Trade-Off Between Bob and Eve}\label{Sec:Sim_tradeoff}

    Fig.~\ref{fig:tradeoff} compares Bob's PSNR against Eve's PSNR. Along the \textbf{LB (swept $\alpha$)} and \textbf{EB (swept $\alpha$)} curves, decreasing $\alpha$ reduces Eve's PSNR but eventually degrades Bob's reconstruction quality as well. \textbf{LB-Unif.} and \textbf{EB-Unif.} correspond to the points that maximize the secrecy rate on their respective curves. \textbf{Adv-DE} lies below both curves, while \textbf{Adv-SE} is omitted due to its substantially lower performance. The \textbf{LB (swept $\alpha$)} curve achieves a higher PSNR at Bob for the same PSNR at Eve and reaches a higher saturation level than the \textbf{EB (swept $\alpha$)} curve. Since the two curves differ in their encoder--decoder pairs trained with unequal and equal information budgets, this result demonstrates the benefit of learning bit-level reliability requirements for improving the PSNR trade-off.

    Fig.~\ref{fig:tradeoff} also shows that all \textbf{LB-Alg.~1} points achieve a lower PSNR at Eve than the \textbf{LB (swept $\alpha$)} curve at the same PSNR at Bob. Since they share the same encoder--decoder pair, this improvement comes from the proposed bit-level power and AN allocation. Among these candidates, \textbf{LB-Alg.~2} selects the profile with the largest $I_{\rm E,min}$, which achieves the highest PSNR at Bob. This selection follows from \textbf{Algorithm~\ref{alg:ProfileSelection}}, which evaluates the trade-off between Bob's reliability and Eve's leakage under the current channel conditions. For the considered SNR setting, the reliability improvement from allocating more power to the information signal outweighs the increased leakage penalty in \eqref{eq:profile_exact_cost}. Therefore, the proposed framework improves the PSNR trade-off through bit-level resource allocation while selecting a suitable target profile according to the channel conditions.

    \subsection{PSNR Performance Versus Angular Separation}\label{Sec:Sim_theta}

      \begin{figure}[t]
        \centering
        {\epsfig{file=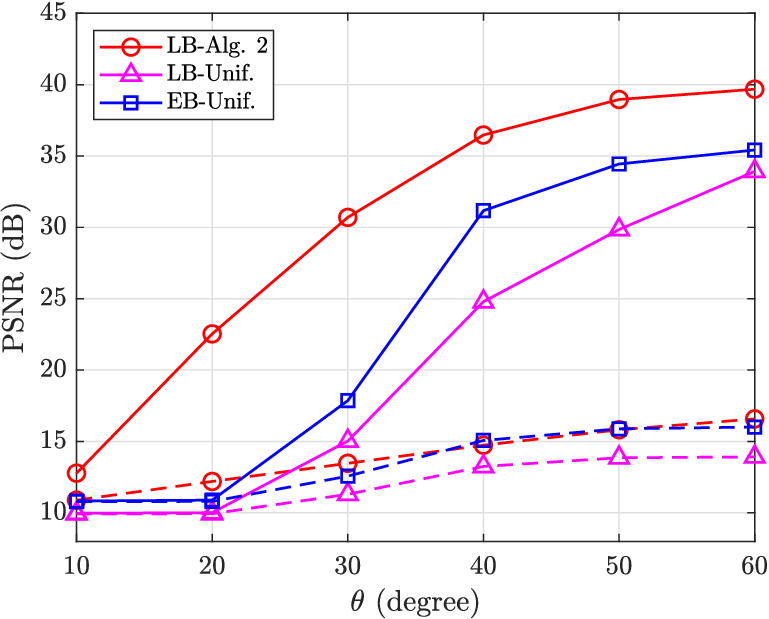,width=0.7\columnwidth}}\vspace{0mm}
        \caption{PSNRs of Bob (solid) and Eve (dashed) versus the minimum angular separation $\theta$ between Bob and Eve for ${\rm SNR}_{\rm B}=12$ dB and ${\rm SNR}_{\rm E}=20$ dB on the CIFAR-10 dataset.}
        \label{fig:psnr_vs_theta}
    \end{figure}

    Fig.~\ref{fig:psnr_vs_theta} compares Bob's and Eve's PSNRs as the minimum angular separation $\theta$ increases. Bob's PSNR improves across all evaluated schemes, with \textbf{LB-Alg.~2} achieving the highest reconstruction quality, while Eve's PSNR remains low and comparable across the schemes. As $\theta$ increases, Eve's worst-case SINR bound decreases, allowing a larger fraction of power to be allocated to the information signal while satisfying the secrecy targets. Furthermore, \textbf{LB-Alg.~2} outperforms \textbf{LB-Unif.} by allocating power according to the learned bit-level reliability and secrecy targets. These results demonstrate that the proposed framework effectively exploits the angular separation to improve Bob's reconstruction quality while maintaining a low reconstruction quality at Eve.


    \subsection{Performance Evaluation in MIMO Systems}\label{Sec:Sim_mimo}

      \begin{figure}[t]
        \centering
        {\epsfig{file=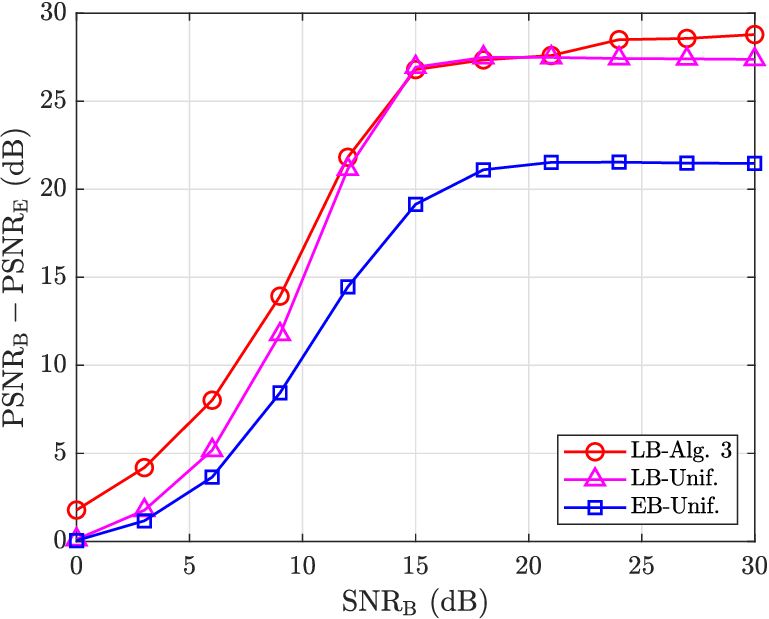,width=0.7\columnwidth}}\vspace{0mm}
        \caption{PSNR gap between Bob and Eve, ${\rm PSNR}_{\rm B}-{\rm PSNR}_{\rm E}$, versus ${\rm SNR}_{\rm B}$ in the MIMO wiretap channel with $N_{\rm tx}=8$, $N_{\rm rx}=N_s=4$, and $\theta_{\rm min}=30^{\circ}$ for ${\rm SNR}_{\rm E}=20$ dB on the CIFAR-10 dataset.}
        \label{fig:psnr_vs_snrb_mimo}
    \end{figure}

    We evaluate the MIMO extension in Sec.~\ref{Sec:MIMO} with $N_{\rm tx}=8$, $N_{\rm rx}=N_s=4$, and $N_z=4$, using the same encoder--decoder pairs and three candidate profiles as in the MISO simulations. The channel matrices $\mathbf{H}_{\rm B}$ and $\mathbf{H}_{\rm E}$ have i.i.d.\ $\mathcal{CN}(0,1/N_{\rm tx})$ entries, and channel pairs violating the minimum angular separation $\theta_{\rm min}=30^{\circ}$ are redrawn. For all schemes, semantic bits are sorted in ascending order of $\mu_{{\rm B},n}$ and divided into four equal groups, which are mapped to Bob's right singular vectors in decreasing order of their singular values. \textbf{LB-Alg.~3} employs \textbf{Algorithm~\ref{alg:MIMO}} with $\lambda=1$ to determine the stream and AN powers using \eqref{eq:MIMO_power_solution} and select a suitable profile for each channel realization.


    Fig.~\ref{fig:psnr_vs_snrb_mimo} compares the PSNR gaps between Bob and Eve over ${\rm SNR}_{\rm B}$ with ${\rm SNR}_{\rm E}=20$ dB. \textbf{LB-Alg.~3} achieves the largest gap among the evaluated schemes, while \textbf{EB-Unif.} exhibits the smallest gap. Unlike uniform allocation, \textbf{LB-Alg.~3} determines the stream powers according to the learned secrecy targets of the bit groups and selects a target profile based on the channel conditions. Meanwhile, the smaller gap of \textbf{EB-Unif.} reflects the limitation of encoder--decoder training with an equal information budget, as discussed in Sec.~\ref{Sec:Sim_tradeoff}. These results demonstrate the effectiveness of the proposed stream-level resource allocation and profile selection for improving the PSNR trade-off in MIMO channels.


    \begin{figure}[t]
        \centering
        {\epsfig{file=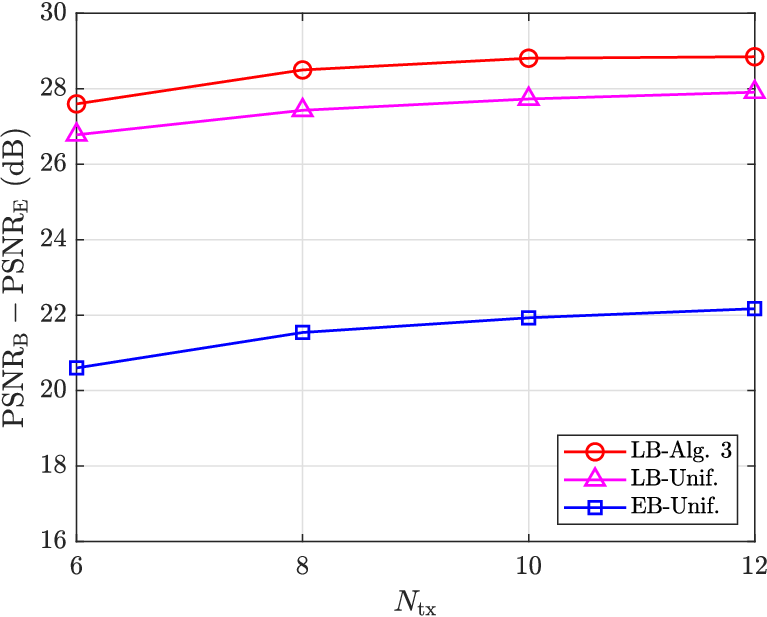,width=0.7\columnwidth}}\vspace{0mm}
        \caption{PSNR gap between Bob and Eve, ${\rm PSNR}_{\rm B}-{\rm PSNR}_{\rm E}$, versus the number of transmit antennas $N_{\rm tx}$ in the MIMO wiretap channel with $N_{\rm rx}=N_s=4$, $\theta_{\rm min}=30^{\circ}$, ${\rm SNR}_{\rm B}=24$ dB, and ${\rm SNR}_{\rm E}=20$ dB on the CIFAR-10 dataset.}
        \label{fig:psnr_gap_vs_antennas_mimo}
    \end{figure}

    Fig.~\ref{fig:psnr_gap_vs_antennas_mimo} compares the PSNR gaps as $N_{\rm tx}$ increases, with $N_{\rm rx}=N_s=4$ and ${\rm SNR}_{\rm B}=24$ dB. The remaining settings are identical to those in Fig.~\ref{fig:psnr_vs_snrb_mimo}. The PSNR gaps increase with $N_{\rm tx}$ for all schemes, which is consistent with the increased spatial degrees of freedom for AN transmission as $N_z=N_{\rm tx}-N_s$ increases. Among the evaluated schemes, \textbf{LB-Alg.~3} consistently achieves the largest PSNR gap, demonstrating the effectiveness of the proposed framework across different transmit antenna configurations.


    \section{Conclusion}\label{Sec:Conclusion}

    In this paper, we proposed a bit-level PLS framework for digital semantic communication using a wiretap SF channel trained in two stages to learn per-bit target bit-flip probabilities for Bob's reliability and Eve's secrecy. For a MISO wiretap channel, we optimized power and AN allocation under worst-case angular separation, deriving a closed-form information-to-AN power ratio and water-filling bit-wise power allocation. We incorporated leakage-aware candidate profile selection and extended the design to MIMO with stream-level closed-form allocation. Simulations showed that leakage concentrates on bits with higher target bit-flip probabilities toward Bob, keeping Eve's reconstruction poor even when Eve decodes most bits more reliably than Bob. By adapting to Bob's channel condition, the framework improved the Bob-Eve reconstruction trade-off and PSNR gap over uniform power allocation and achieved higher reconstruction quality at Bob than the evaluated adversarial learning-based schemes.


    An important direction for future research is to evaluate and enhance the proposed framework against an adaptive eavesdropper that trains its own decoder, beyond the shared-decoder model considered in this paper. Another promising direction is to develop a rigorous worst-case secrecy analysis for MIMO wiretap channels that does not rely on the diagonal approximation of the subspace overlap. Integrating the proposed framework with channel coding, so that the learned reliability and secrecy targets can be realized jointly with unequal error protection, would also be an important direction for future research.

    \appendices
    \section{Proof of Proposition 1}\label{Apdx:Prop1}
    The Lagrangian of \textbf{(P2)} is
    \begin{align}
        \mathcal{L}({\bar p_n},\nu) = \sum_{n=1}^N \omega_{n} \exp(-A\alpha_n^\star \bar p_n)+\nu\left(\sum_{n=1}^N \bar p_n - P_{\rm tot}\right),
    \end{align}
    where $\nu$ is the Lagrange multiplier associated with the power budget constraint. The stationarity condition $\partial \mathcal{L}/\partial \bar{p}_n = 0$ gives $-A\alpha_n^\star \omega_{n} \exp(-A\alpha_n^\star \bar{p}_n) + \nu = 0$, and so
    \begin{align}
         \bar{p}_n = \frac{1}{A\alpha_n^\star}\ln\!\left(\frac{A\alpha_n^\star\omega_{n}}{\nu}\right).
    \end{align}
    Since the objective in \textbf{(P2)} is convex in $\bar{p}_n$ and the constraint set is convex, this stationary point is the unique minimizer whenever it is non-negative. Incorporating the non-negativity constraint $\bar{p}_n\geq 0$ via the complementary slackness condition of the KKT system then yields the water-filling solution in \eqref{eq:water_filling}, with $\nu$ set such that $\sum_{n=1}^N \bar p_n^\star = P_{\rm tot}$. $\blacksquare$

    \bibliographystyle{IEEEtran}
    \bibliography{Reference}
    
\end{document}